\documentclass[longauth]{aa}

\usepackage{graphicx}

\usepackage{placeins}           

\usepackage[varg]{txfonts}
\usepackage{upgreek} 
\usepackage{natbib}

\usepackage{hyperref}

\hypersetup{%
  colorlinks=true,
  linkcolor=blue,
  citecolor=blue
}

\newcommand{\CI}{C\,{\sc i}}

\newcommand{\CII}{C\,{\sc ii}}
\newcommand{\HII}{H\,{\sc ii}}

\begin{document} 


\title{PDRs4All}
\subtitle{XII. Far-ultraviolet-driven formation of simple hydrocarbon radicals and their relation with polycyclic aromatic hydrocarbons}

\titlerunning{FUV-driven formation of simple hydrocarbon radicals and their relation with PAHs} 
\authorrunning{Goicoechea et al.}

 \author{J. R.\,Goicoechea\inst{1}
          \and
        J. Pety\inst{2,3}
        \and
        S. Cuadrado\inst{1}
        \and
        O. Bern\'e\inst{4}
        \and    
        E. Dartois\inst{5}    
        \and    
        M. Gerin\inst{3}   
        \and    
        C. Joblin\inst{4}    
        \and    
        J. K\l{}os\inst{6,7}    
        \and    
        F. Lique \inst{8}
        \and    
        \mbox{T. Onaka} \inst{9}    
        \and    
        \mbox{E. Peeters}\inst{10,11,12}
        \and   
        \mbox{A. G. G. M. Tielens} \inst{13,14}    
        \and    
        \mbox{F. Alarc\'on} \inst{15}   
        \and    
        \mbox{E. Bron} \inst{16}    
        \and    
        \mbox{J. Cami}\inst{10,11,12}
        \and    
        \mbox{A. Canin}\inst{4}
        \and    
        \mbox{E. Chapillon}\inst{17,2}
        \and    
        \mbox{R. Chown}\inst{10,11,18}
        \and    
        \mbox{A. Fuente}\inst{19}
        \and    
        \mbox{E. Habart} \inst{20}    
        \and
        \mbox{O. Kannavou} \inst{20}    
        \and  
        \mbox{F. Le Petit} \inst{16}     
        \and    
        \mbox{M. G. Santa-Maria}\inst{21,1}
        \and    
        \mbox{I. Schroetter }\inst{4}    
        \and    
        \mbox{A. Sidhu}\inst{10,11}
        \and    
        \mbox{B. Trahin} \inst{20}    
        \and
        \mbox{D. Van De Putte}\inst{10,11}
        \and    
        \mbox{M. Zannese} \inst{20}    
      }

\institute{Instituto de F\'{\i}sica Fundamental
     (CSIC). Calle Serrano 121-123, 28006, Madrid, Spain. \email{javier.r.goicoechea@csic.es}
\and
 Institut de Radioastronomie Millim\'etrique, 38406, 
 Saint Martin d’H$\grave{\rm e}$res, France.
\and
LUX, Observatoire de Paris, Universit\'e PSL, Sorbonne Universit\'e, CNRS, 75014, Paris, France.  
\and
Institut de Recherche en Astrophysique et Plan\'etologie, Universit\'e de Toulouse, CNRS, CNES, Toulouse, France.
\and
Institut des Sciences Mol\'eculaires d’Orsay, CNRS, Universit\'e Paris-Saclay, Orsay, France.
\and
Joint Quantum Institute, Department of Physics, University of Maryland, College Park, MD 20742, USA.
\and
Dept. of Physics, Temple University, Philadelphia, PA 19122, USA.
\and
Univ. Rennes, CNRS, IPR (Institut de Physique de Rennes), UMR6251, 35000, Rennes, France.
\and
Department of Astronomy, Graduate School of Science, The University of Tokyo, 7-3-1 Bunkyo-ku, Tokyo 113-0033, Japan.
\and
Department of Physics and Astronomy, University of Western Ontario, London, Ontario, Canada.
\and
Institute for Earth and Space Exploration, University of Western Ontario, London, Ontario, Canada.
\and
Carl Sagan Center, SETI Institute, Mountain View, CA, USA. 
\and
Leiden Observatory, Leiden University, Leiden,  The Netherlands.
\and
Astronomy Dept., University of Maryland, College Park, MD, USA.
\and
Dipartimento di Fisica, Università degli Studi di Milano, Via Celoria 16, 20133 Milano, Italy.
\and
LUX, Observatoire de Paris, Universit\'e PSL, Sorbonne Universit\'e, CNRS, 92190, Meudon, France.
\and
Laboratoire d'Astrophysique de Bordeaux, Universit\'e de Bordeaux, CNRS, F-33615 Pessac, France.
\and
Department of Astronomy, The Ohio State University, 140 West 18th Avenue, Columbus, OH 43210, USA.
\and
Centro de Astrobiolog\'{i}a (CAB), CSIC-INTA, Ctra. de Torrej\'on a Ajalvir, km 4, 28850 Torrej\'on de Ardoz, Spain.
\and
Universit\'e Paris-Saclay, CNRS, Institut d’Astrophysique Spatiale, Orsay, France.
\and
Department of Astronomy, University of Florida, P.O. Box 112055, Gainesville, FL 32611, USA.}

   \date{Received 9 December 2024 / Accepted 5 March 2025}

\abstract 
{The infrared emission from  polycyclic aromatic hydrocarbons (PAHs), along with emission from atomic carbon and simple hydrocarbons, is a robust tracer of the
interaction between stellar far-UV (FUV) radiation and molecular clouds.
We present subarcsecond-resolution ALMA mosaics  of the Orion Bar photodissociation region (PDR) in  [\CI]\,609\,$\upmu$m ($^3$P$_1$--$^3$P$_0$), \mbox{C$_2$H ($N$\,=\,4--3)}, and \mbox{C$^{18}$O ($J$\,=\,3--2)} emission lines 
complemented by JWST images of H$_2$ and aromatic infrared band (AIB) emission.
We interpreted the data using up-to-date PDR and radiative transfer models, including 
high-temperature \mbox{C$_2$H\,(X$^2\Sigma^+$)--$o$/$p$-H$_2$}
and \mbox{C\,($^3$P)--$o$/$p$-H$_2$} inelastic collision rate coefficients
(we computed the latter  up to 3000~K).
The rim of the Bar shows very corrugated and filamentary structures made of small-scale H$_2$ dissociation fronts (DFs). 
The [\CI]\,609\,$\upmu$m emission peaks very close ($\lesssim$\,0.002\,pc) to the main H$_2$-emitting DFs, suggesting the presence of gas density gradients. 
These DFs are also bright and remarkably similar in \mbox{C$_2$H} emission, which traces "hydrocarbon radical peaks" characterized by very high C$_2$H  abundances, reaching up to several $\times$10$^{-7}$.
The high abundance of C$_2$H and of related hydrocarbon radicals, such as CH$_3$, CH$_2$, and CH, can be attributed to gas-phase reactions driven by elevated temperatures, the presence of C$^+$ and C,  and the  reactivity of \mbox{FUV-pumped H$_2$}. 
The hydrocarbon radical peaks roughly coincide with maxima of the 3.4/3.3\,$\upmu$m AIB  intensity ratio, which is a proxy for the 
aliphatic-to-aromatic content of PAHs.
This implies that the conditions triggering the formation of simple hydrocarbons also favor the formation (and survival) of PAHs with aliphatic side groups, potentially via 
the contribution of bottom-up processes in which abundant hydrocarbon radicals react in situ with PAHs.
Ahead of the DFs, in the  atomic PDR zone (where \mbox{[H]\,$\gg$\,[H$_2$]}),
the AIB emission is the brightest, but small PAHs and carbonaceous grains
undergo photo-processing due to the stronger FUV field. Our detection of trace  amounts of C$_2$H in this zone may result from the photoerosion of these species.
This study provides a spatially resolved view of the chemical stratification of key carbon carriers in a PDR. 
Overall, both bottom-up and top-down processes appear to link simple hydrocarbon molecules with PAHs in molecular clouds; however, the exact chemical pathways and their
relative contributions remain to be quantified.
}

\keywords{ISM: abundances -- ISM: Molecules, Molecular data, Molecular processes --- photon-dominated region (PDR)}
\maketitle
%

\section{Introduction}\label{sec:introduction}

Far-UV (FUV) radiation \mbox{\mbox{(6 $<$\,$E$\,$<$13.6\,eV})} from massive stars  \mbox{penetrates} their natal giant molecular clouds \citep[e.g.,][]{Goicoechea15,Miriam23}, heating their gas and dust, altering their chemistry, and driving flows that photoevaporate the cloud \citep[e.g.,][]{Maillard21}.
The interaction of FUV radiation and \mbox{interstellar} matter occurs in "\mbox{photodissociation regions}" \citep[PDRs; e.g.,][]{Wolfire22} where FUV \mbox{radiation} 
\mbox{regulates} the  heating, ionization, and chemistry. 
PDRs release bright infrared (IR) emission, in particularl both \mbox{collisionally} excited and \mbox{FUV-pumped} molecular hydrogen (H$_2$) line emission and aromatic infrared band emission (AIB),
 which mainly results from IR fluorescence of \mbox{FUV-pumped}
polycyclic aromatic hydrocarbons \citep[PAHs,][]{Leger84,Allamandola85}. The PAHs in PDRs lock up as much as 10\%~of interstellar carbon \citep[e.g.,][]{Tielens08}.

\begin{figure*}[th]
\centering   
\includegraphics[scale=0.332, angle=0]{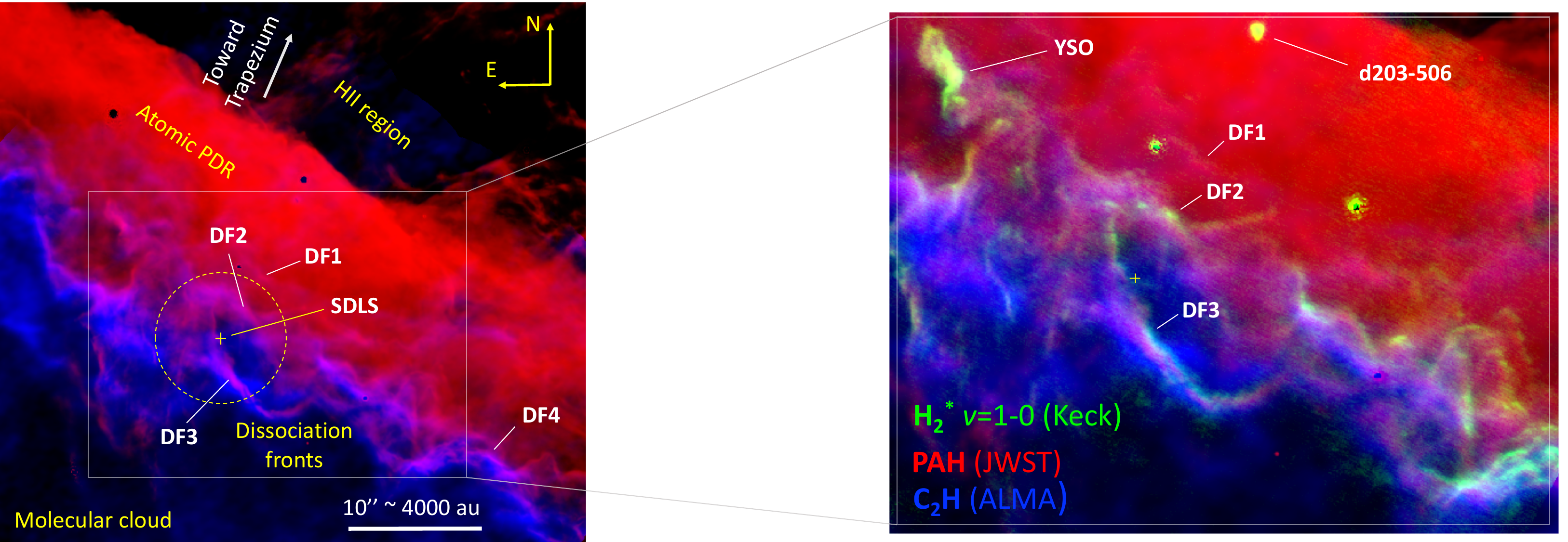}
\caption{Subarcsecond view of the Orion Bar PDR. 
 \textit{Left:} ALMA \mbox{C$_2$H $N$\,=4--3}  line emission (blue) and \mbox{JWST  F335M$-$F330M image} (red), a proxy of the $\sim$3.3\,$\upmu$m AIB feature \citep{Habart24}. 
 For reference, we show the IRAM\,30\,m SDLS at 
\mbox{$\alpha$(2000)\,=\,5:35:20.8} and \mbox{$\delta$(2000)\,=\,$-$05:25:17} with a dashed circle centered at the yellow cross.
 \textit{Right:} Zoom-in of a smaller field adding a Keck \mbox{H$_{2}$ $v$\,=\,1--0 $S$(1)} image (green) from \cite{Habart23a}.
Table~\ref{table:coordinates} shows the coordinates of the main DFs  discussed in this work.}
\label{fig:RGB_Bar}
\end{figure*}

 Carbon  is an abundant cosmic element \mbox{\citep[][]{Asplund09}}. Due to its allotropy, simple hydrocarbon molecules are the \mbox{building} blocks of carbon chemistry in space. They participate in the formation of  complex organic molecules, PAHs, and carbonaceous grains \mbox{\citep[][]{Jones12}}. 
With an ionization potential of 11.3\,eV (lower than that of H), 
carbon ions (C$^+$) play a pivotal role in  \mbox{cooling}  \mbox{FUV-irradiated} neutral 
gas via the \mbox{[\CII]\,158\,$\upmu$m}  line \mbox{\citep[][]{Dalgarno72}}. 
  \mbox{Additionally}, C$^+$ initiates the carbon chemistry of FUV-irradiated H$_2$ gas
\citep[][]{Stecher72,Freeman82,Tielens85,Sternberg95,Agundez10},  triggering the formation of carbon hydrides \mbox{\citep[e.g.,][]{Gerin16}}.
Dust extinction and gas absorption progressively reduce the flux and energy of FUV photons inside PDRs \citep[][]{Flannery80,Goico07}. At a given cloud depth, depending on FUV flux ($G_0$) and  gas density ($n_{\rm H}$), the dominant gas-phase carbon reservoir transitions
from C$^+$ to C to CO. These key transformations take place in the so-called  \mbox{CO\,/\,C\,/\,C$^+$} transition zone of a PDR \citep[e.g.,][]{Tielens85,Sternberg95,Goico07}.

The AIB emission at 3.3, 6.2, 7.7, 8.6, 11.2, and 12.7\,$\upmu$m is a specific tracer of PDR environments and dominates the IR spectra of many galactic sources
and star-forming galaxies \mbox{\citep[][]{Tielens08}}.
The observed changes in AIB emission in these sources suggest that the species responsible for this emission undergo changes and photochemical modifications
\mbox{\citep[][]{Peeters02}}. Much of the PAH photoprocessing is controlled by the
ratio of the FUV flux to H atom density, $G_0$/$n$(H) \cite[][]{Montillaud13,Andrews16}.
Observational evidence has shown that close to PDR edges, the smallest PAHs are destroyed and the aliphatic content of the AIB carriers is reduced 
\citep[e.g.,][]{Joblin96,Peeters24,Chown24}.

The general understanding of the formation, destruction, and reactivity of PAHs is not settled and is rapidly evolving. Some studies have suggested that PAHs form through the photoprocessing of very small carbonaceous grains \citep[e.g.,][]{Cesarsky00,Berne07,Pilleri12}.
This is a \mbox{"top-down"} mechanism. In addition,
the recent radio detection in \mbox{TMC-1} of polar cyano derivatives of small PAHs 
\citep[][]{McGuire18,McGuire21,Wenzel24a,Wenzel24b,Cernicharo24}
and the detection of indene \citep[\mbox{$c$-C$_9$H$_8$},  a very simple asymmetric  PAH;][]{Cernicharo21}  
demonstrates that free-flying, gas-phase PAHs exist in dark molecular clouds before star formation \mbox{(i.e., before FUV processing)}. This suggests that
PAHs already existed and/or that 
ion-molecule reactions and radical chemistry 
\cite[e.g.,][]{Kaiser15,Zhao19,He20,Lemmens22,Levey22} contribute to the in situ growth of PAHs in cold  (\mbox{$\simeq$10--20\,K}) gas.
The much higher temperatures in PDRs (up to several hundred Kelvin), along with the enhanced abundances of  simple hydrocarbons triggered by FUV-induced chemistry \citep[starting with
 \mbox{CH$^+$};][]{Nagy13,Joblin18,Goico19}, may provide another  environment for \mbox{"bottom-up"} PAH formation that is followed by PAH photoprocessing.
 
Observations of interstellar PDRs have revealed
open-shell\footnote{We give the ground electronic state in parentheses.} hydrocarbon radicals such as 
\mbox{CH ($^2$$\Pi$)},
\mbox{CH$_2$ ($^3$B$_1$)}, \mbox{C$_2$H ($^2$$\Sigma^+$)}, \mbox{$c$-C$_3$H ($^2$B$_2$)}, \mbox{$l$-C$_3$H ($^2$$\Pi$)}, and \mbox{C$_4$H ($^2$$\Pi$)}, which are highly  reactive,
as well as closed-shell hydrocarbons, such as \mbox{CH$^+$ ($^1$$\Sigma^+$)},
\mbox{$c$-C$_3$H$_2$ ($^1$A$_1$)}, \mbox{$l$-H$_2$C$_3$ ($^1$A$_1$)}, 
\mbox{$l$-C$_{3}$H$^+$ ($^1$$\Sigma^+$)}, and \mbox{CH$_{3}^{+}$ ($\tilde{X}^1$A$'$)}
\citep[][]{Fuente03,Teyssier04,Polehampton05,Pety05,Pety12,Pilleri13,Guzman15,Cuadrado15,Tiwari19,Nagy15,Nagy17,Goico19,Zannese25}.
The origin of this rich  chemistry in such harsh environments is still debated.

Early gas-phase PDR models underestimated the observed abundances  
of some of these hydrocarbons, especially as the number of carbon atoms increases. 
In addition, at the moderate angular resolution provided by the previous generation of telescopes, the AIB  and hydrocarbon emission peaks seemed to approximately coincide. This led astronomers to suggest that the high abundances of specific hydrocarbons 
(which we generally denote as C$_x$H$_y$) could result from the photodestruction of PAHs or very small carbonaceous grains  \citep[][]{Fosse00,Fuente03,Teyssier04,Pety05,Alata15,Guzman15}, and there is experimental support for such a link \citep{Jochims94,LePage03,Allain96,Joblin03,Bierbaum11,Alata14,Marciniak21,Tajuelo24}.
However, gas-phase production of hydrocarbons in a PDR may also be linked to the high
temperatures and the enhanced reactivity of FUV-pumped H$_2$  \citep{Cuadrado15}.
With JWST and ALMA, we can spatially resolve the emission from H$_2$, PAHs, and simple hydrocarbons, allowing us to accurately link it to the steep physicochemical gradients in PDRs. This capability can further elucidate these opposing yet perhaps complementary possibilities as well as the processes of PAH growth and destruction.
This is also a relevant question in the extragalactic context, where the abundance of simple hydrocarbons such as C$_2$H can be  high in massive star-forming environments exposed to strong irradiation \citep[e.g.,][]{Meier15,Burillo17}.

In this paper we target the Orion Bar, the prototypical strongly irradiated PDR  \mbox{(Fig.~\ref{fig:RGB_Bar})}. 
In \mbox{Sect.~\ref{sec:observations}}, we introduce the source and observational dataset.
In \mbox{Sect.~\ref{sec:results}}, we discuss the most salient features of the ALMA and JWST spectroscopic-images taken
by \citet{Berne22}. 
In \mbox{Sect.~\ref{sec:analysis}}, we analyze these images with emission crosscuts as well as nonlocal thermodynamic equilibrium (LTE) radiative transfer models.
In \mbox{Sect.~\ref{sec:pdr_mods}}, we model the chemistry of  C$_2$H and related radicals. Finally, in \mbox{Sect.~\ref{sec:discussion}}, we
 try to find links with the observed AIB emission.

 \section{Observations}\label{sec:observations}
\subsection{The Orion Bar photodissociation region}

  The Orion Bar is an interface of the Orion molecular cloud (\mbox{OMC-1}) and the  \mbox{Huygens} \HII~region (M42) photoionized by massive stars in the Orion Nebula cluster, mainly  \mbox{$\theta^1$ Ori C} in the Trapezium  \citep[e.g.,][]{ODell01}. 
 At an adopted distance of 414~pc \citep[see dicussion in ][]{Habart24},
this region is the closest cluster environment with ongoing high-mass star-formation. The  Bar is a strongly irradiated PDR \citep[with $G_0$\,$>$\,10$^4$,][]{Marconi98,Peeters24}.   
The ionization front (IF) at the edge of the Bar marks the transition between the   \HII~region (with \mbox{[H$^+$]\,$\gg$\,[H]}, where [X] refers to the abundance of species X relative to H nuclei) and the neutral and predominantly atomic-gas zone of the PDR, 
where \mbox{[H]\,$\gg$\,[H$_2$]} and 
\mbox{$n$(H)\,$\simeq$\,(5--10)$\times$10$^4$\,cm$^{-3}$}
\citep[e.g.,][]{Tielens93,Pellegrini09,vanderWerf13,Habart24,Peeters24}. 
At \mbox{$\sim$10-20$''$} \mbox{($\sim$0.02-0.04\,pc)} from the IF, the FUV flux is  attenuated enough that
\mbox{[H$_2$]\,$\gg$\,[H]}. This is  the \mbox{H$_2$\,/\,H} transition zone, or dissociation front (DF), where most of the gas becomes molecular and some structures reach up to 
\mbox{$n_{\rm H}$\,=\,$n$(H)\,+\,2$n$(H$_2$)\,$\gtrsim$\,10$^6$\,cm$^{-3}$}  \citep[e.g.,][]{Goico16}. The  DF  displays a plethora of  ro-vibrational  H$_2$ lines 
\citep{Parmar91,Luhman94,Allers05,Shaw09,Kaplan21,Peeters24,Putte24}
and AIB emission \citep{Sloan97,Chown24}. In the standard view
of a constant-density PDR,  the \mbox{CO\,/\,C\,/\,C$^+$} transition zone is expected to occur beyond the DF \mbox{\citep[e.g.,][]{Tielens85}}. However, owing to low-angular resolution, [\CII]\,158\,$\upmu$m 
and \mbox{[\CI]\,370,609\,$\upmu$m} observations have not accurately settled the exact position of this zone
\citep[e.g.,][]{Tauber95,Wyrowski97,Ossenkopf13,Goicoechea15,Cuadrado19}.

\subsection{ALMA imaging}\label{sec:alma-obs}

We carried out a $\sim$40$''$$\times$40$''$ mosaic of the Bar \mbox{(Fig.~\ref{fig:RGB_Bar})} using forty-seven ALMA \mbox{12 m} antennas at the frequency of the 
\mbox{C$_2$H~$N$\,=\,4--3} (349.3\,GHz\footnote{We refer to the  C$_2$H
$F$\,=\,5--4 and $F$\,=\,4--3 hyperfine structure (HFS) lines
of the $N$\,=\,4--3, $J$\,=\,9/2--7/2 fine-structure transition. These two  HFS lines overlap and result in a single, unresolved line at $\sim$349.339\,GHz.
Due to this blending, the resulting line profile may not provide precise line widths if line overlap and opacity  effects are significant 
\citep[different HFS lines can have different excitation temperatures, opacities, and  widths; for HCN HFS lines, see,][]{Goico22}.} \mbox{$N$\,=\,4--3}), \mbox{C$^{18}$O~$J$\,=\,3--2} (329.3\,GHz)  and [\CI]\,609\,$\upmu$m lines (in three different frequency setups).  
These  observations belong to project  \mbox{2021.1.01369.S} (\mbox{P.I.: J. R. Goicoechea})
 and consisted of a 27-pointing (C$_2$H and C$^{18}$O) and 52-pointing ([\CI]\,609\,$\upmu$m) mosaics centered at  \mbox{$\alpha$(2000) = 5$^h$35$^m$20.6$^s$}; 
\mbox{$\delta$(2000) = $-$05$^o$25$'$20.0$''$}. We observed the \mbox{C$_2$H~$N$\,=\,4--3} and  \mbox{C$^{18}$O~$J$\,=\,3-2} lines (band~7),  as well as the [\CI]\,609\,$\upmu$m line (band~8) 
using  correlators providing $\sim$282\,kHz and $\sim$564\,kHz resolution, respectively. 
We binned all spectra to a common velocity resolution of 0.4\,km\,s$^{-1}$.
The total
observation times with the ALMA\,12\,m array 
were $\sim$2.7\,h and 4.6\,h, respectively. In order to recover the extended emission
(a few tens of arcseconds) filtered out by the interferometer, we used 
ACA\,7\,m array,  as well as  fully sampled single-dish maps obtained with the total-power (TP) antennas as zero- and short-spacings.

We calibrated the interferometer data using the standard ALMA pipeline in
CASA.  We calibrated  the TP data and converted it into a
position-position-velocity cube also using CASA. Following~\citet{rodriguez08}, we used the \texttt{GILDAS/MAPPING} software
to create  the short-spacing
visibilities not sampled by the ALMA-7m and ALMA-12m interferometers:

(i) We first use the TP and ALMA-7m to produce a clean image containing
all spatial information from 0 to 30\,m, that is, equivalent to a total
power telescope of 30\,m-diameter. In this step, the TP map is
deconvolved from the TP beam in the Fourier plane before
multiplication by the ALMA-7m primary beam in the image plane. After a
last Fourier transform, we sampled the pseudo-visibilities between 0 and
5\,m, the difference between the diameters of the TP and the
ALMA-7m antennas. These visibilities were then merged with the ALMA-7m
interferometric observations. Each mosaic field was imaged and a dirty
mosaic was built combining those fields in the following optimal way in
terms of signal--to--noise ratio~\citep{pety10}. The dirty image was
deconvolved using the standard H\"ogbom CLEAN algorithm. The clean image
has an elliptical beam that is rounded by smoothing with a Gaussian to
the angular resolution corresponding to a  telescope of 30\,m.

\begin{figure*}[h]
\centering   
\includegraphics[scale=0.56,angle=0]{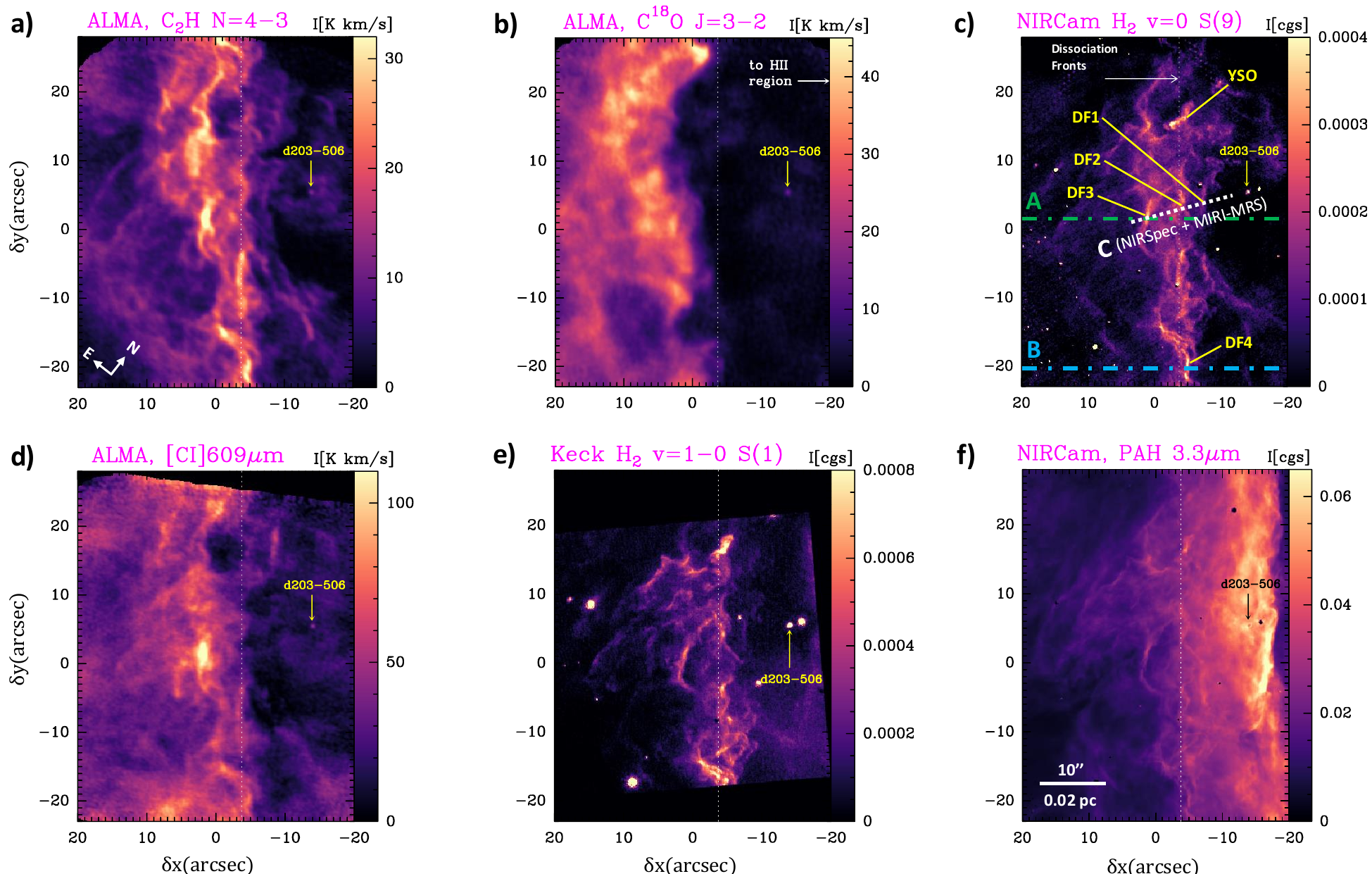}
\caption{ALMA, JWST/NIRCam, and Keck images of the Orion Bar PDR.
We rotated all images to bring the FUV-illuminating direction from the Trapezium stars into the horizontal direction and from the right.
The center of all images is at
\mbox{$\alpha$(2000) = 5$^h$35$^m$20.50$^s$}; 
 \mbox{$\delta$(2000) = $-$05$^o$25$'$21.4$''$}.
 The vertical dashed white
line marks the approximate position of the main H$_2$ DFs parallel to the Bar. \mbox{(a) \mbox{C$_2$H $N$\,=4--3}} and
\mbox{(b) \mbox{C$^{18}$O $J$\,=3--2}}.
\mbox{(c) NIRCam F470M$-$F480M image}, a proxy of the H$_2$ $v$\,=\,0 $S$(9) line at  4.69\,$\upmu$m \citep{Habart24}. Dashed and dotted lines show the position
and orientation of the intensity crosscuts A, B (see Fig.~\ref{fig:crosscuts}), 
and C (the JWST spectroscopy cut, Fig.~\ref{fig:PDR_MIRI_crosscut}) discussed in the text.
\mbox{(d) \mbox{[\CI]\,\,609\,$\upmu$m}. 
\mbox{(e) Keck/NIRC2} image around the \mbox{H$_2$ $v$\,=\,1--0 $S$(1)} line at 2.2\,$\upmu$m \citep{Habart23a}.
\mbox{(f)} NIRCam  F335MR$-$F330M image}, a proxy of the AIB emission at  3.3\,$\upmu$m \citep{Habart24}. An arrow marks  the position of protoplanetary disk d203-506. The units
``cgs'' refer to \text{erg\,s$^{-1}$\,cm$^{-2}$\,sr$^{-1}$}.
We show more details in the combined RGB images in Fig.~\ref{fig:contours}.}
\label{fig:cch_c18o_jwst_integrated_intensity}
\end{figure*}

(ii) We then used the resulting TP+ALMA-7m image with a typical resolution of
$6''$ to  create the short-spacing visibilities not sampled by the
ALMA-12m interferometer. The TP+ALMA-7m map is deconvolved from the
TP+ALMA-7m beam in the Fourier plane before multiplication by the
ALMA-12m primary beam in the image plane. After a last Fourier
transform, pseudo-visibilities were sampled between 0 and 12\,m, the
ALMA-12m diameter. In principle, we could have sampled between 0 and
18\,m. In practice, the ALMA-12m visibilities above 12\,m have a
better quality that the one obtained from TP+ALMA-7m map above
12\,m. These visibilities were then merged with the ALMA-12m
interferometric observations and the resulting dataset was deconvolved
as in step 1. The resulting data cube 
\mbox{(Fig.~\ref{fig:cch_c18o_jwst_integrated_intensity})} were then scaled from Jy\,beam$^{-1}$ flux units to
main beam temperature, $T_{\rm mb}$,  scale using the synthesized beam size.

The final synthesized beams are \mbox{0.52$''$\,$\times$\,0.38$''$}
at position angle \mbox{PA\,=\,110$^{\circ}$} (492\,GHz),
\mbox{0.78$''$\,$\times$\,0.50$''$} at
at \mbox{PA\,=\,48$^{\circ}$} (349.3\,GHz), and
\mbox{0.77$''$\,$\times$\,0.60$''$} at  \mbox{PA\,=\,64$^{\circ}$} (329.3\,GHz). This implies nearly a factor  $\sim$30 better resolution than previous \mbox{C$_2$H~$N$\,=\,4-3} and [\CI]\,609\,$\upmu$m single-dish maps of the Orion Bar \citep{Wiel09,Tauber95}.
The typical rms noise of the final cubes are $\sim$\,0.19\,K (at 349.3\,GHz),
$\sim$\,0.24\,K (at 329.3\,GHz), and $\sim$\,0.63\,K (at 492\,GHz)
per velocity channel.

\subsection{JWST and Keck infrared observations: PAHs and H$_2$}\label{sec:jwst-obs}  
  
We use JWST IR images of the H$_2$  and AIB emission obtained and calibrated by  \citet{Berne22}.
Here we analyze  JWST/NIRCam   \mbox{F335M--F330M} and \mbox{F470M--F480M}
(continuum subtracted) photometric images  as proxies of
the \,3.3\,$\upmu$m AIB  and \mbox{H$_2$ $v$\,=\,0--0 $S(9)$} emission, respectively \citep[][]{Habart24}. The point-spread-function (PSF) FWHM of these images is \mbox{$\sim$0.1--0.2$''$}. We also make use of the narrow-band filter
 image at 2.12\,$\upmu$m, dominated by the  H$_2$ $v$\,=\,1--0 $S(1)$ line emission,
 obtained by us  using
Keck/NIRC2 adaptive optics at $\sim$0.1$''$ resolution \citep{Habart23a}.  
Figure~\ref{fig:RGB_Bar} shows a composite RGB image using these images.
We further employ near-IR NIRSpec \citep{Peeters24} and
mid-IR MIRI-MRS mosaics \citep{Chown24,Putte24}. The field of view (FoV) of these spectroscopic-images is notably smaller, about \mbox{27$''$\,$\times$\,6$''$}, but include the main DFs and the externally irradiated protoplanetary disk \mbox{d203-506}, observed
by us with ALMA and JWST spectrometers
\citep[e.g.,][]{Berne23,Berne24,Goico24}.
In addition, we utilize images of the AIB 3.4\,$\upmu$m over 3.3\,$\upmu$m  band ratio inferred with NIRSpec at about 0.1$''$ resolution \citep{Peeters24}.
Finally, we use \mbox{MIRI-MRS} images of the low-energy 
 H$_2$ \mbox{$v$\,=\,0--0} pure rotational emission lines  $S$(1), $S$(2), $S$(3), and $S$(4) at 17.03, 12.28\,$\upmu$m, 9.66\,$\upmu$m, and  8.02\,$\upmu$m respectively.
The PSF FHWM of these observations varies between  0.7$''$ and 0.3$''$.

\subsection{IRAM\,30\,m and Herschel/HIFI multi-$N$ C$_2$H observations}\label{sec:survey-obs}

To get a broader view of the Bar and to study beam-dilution effects of 
single-dish  observations, we  obtained  a \mbox{2.5$'$\,$\times$\,2.5$'$}  
map of the \mbox{C$_2$H $N$\,=\,4--3} emission over the entire region (see Fig.~\ref{fig:map_e330}), using the IRAM\,30\,m telescope (Pico Veleta, Spain) with the  E330 receiver and the FTS backend at 200\,kHz  resolution. We carried out \mbox{on-the-fly}  scans  along and perpendicular to the Bar. The resulting spectra were gridded to a data cube through convolution with a Gaussian kernel providing a final resolution of $\sim$8$''$.
The total integration time was $\sim$6\,h during excellent winter conditions
($\lesssim$\,1\,mm of precipitable water vapor). The achieved rms noise is $\sim$1\,K per resolution channel.

To better understand the excitation of C$_2$H  lines and to 
accurately constrain our PDR models, we complemented our analysis with existing observations of multiple C$_2$H  rotational lines ($N$\,=\,1--0 to 4--3) obtained with the IRAM\,30\,m telescope  ($\sim$8$''$ to $\sim$28$''$ resolution) toward 
 the ``single-dish line survey'' position (SDLS; see \mbox{Fig.~\ref{fig:RGB_Bar} left}) that includes DFs observed with ALMA and JWST,
at \mbox{$\mathrm{\alpha_{2000}=05^{h}\,35^{m}\,20.8^{s}\,}$}, 
\mbox{$\mathrm{\delta_{2000}=-\,05^{\circ}25'17.0''}$}. 
 \cite{Cuadrado15} first presented this data.
In addition, we complemented our dataset by including rotationally excited C$_2$H lines
 ($N$\,=\,6--5 to 
$N$\,=\,10--9) detected by \cite{Nagy15,Nagy17} with the Herschel Space Observatory
 toward the   \mbox{``CO$^+$ peak''} position \mbox{\citep{Stoerzer95}}.
This position is located at only  $\sim$4$''$ from the SDLS position. Thus, within the
area subtended by the Herschel beams.
  These observations were carried out with HIFI  \citep{deGra10} at a spectral-resolution of 1.1\,MHz (0.7\,km\,s$^{-1}$ at 500\,GHz). HIFI's 
 angular resolution ranges from  $\sim$20$''$ to $\sim$42$''$ \citep[][]{Roelfsema12}.

Because the beam size of the IRAM\,30\,m and Herschel  telescopes varies with frequency, the
observation of  multiple  C$_2$H rotational lines provides the line intensity  averaged over slightly different areas of the Bar. These  observations do not spatially resolve the  emission arising from the 
different small-scale DFs. To approximately  correct for these beam-size differences, we estimated a frequency-dependent \mbox{`beam coupling factor'} ($f_{\rm b}$)    using the spatial information provided by the high-angular resolution \mbox{C$_2$H $N$\,=\,4--3} map taken with ALMA
(see \mbox{Appendix~\ref{App:beam_dilution}}).

\section{Results}\label{sec:results}

Figure~\ref{fig:RGB_Bar} (right panel) shows a RGB image composed of JWST/NIRCam \mbox{(red, AIB\,3.3\,$\upmu$m; the C--H stretching mode)}, Keck (green, \mbox{H$_2$ $v$\,=\,1--0 $S$(1)}) and ALMA (blue, \mbox{C$_2$H~$N$\,=\,4--3})  observations of the southern edge of the Orion Bar. Labels indicate the main structures and objects  in this FoV. The RGB image reveals a very structured DF made of small-scale fronts \citep[see][]{Habart24}.  These fronts are engulfed by a PAH-emitting \mbox{"halo"} that separates the predominantly neutral atomic gas edge of the cloud from the adjacent \HII~region
(hot ionized gas).
Figure~\ref{fig:cch_c18o_jwst_integrated_intensity}  shows 
\mbox{$\sim$40$''$\,$\times$\,40$''$} (\mbox{$\sim$0.08\,pc\,$\times$\,0.08\,pc})
  images of multiple tracers individually:
(a)~C$_2$H $N$\,=\,4--3, (b)~\mbox{C$^{18}$O $J$\,=\,3--2},
(c)~\mbox{H$_2$ $v$\,=\,0--0 $S(9)$},
(d)~\mbox{[\CI]\,\,609\,$\upmu$m}; (e) \mbox{H$_2$ $v$\,=\,1--0 $S(1)$};
and (f)~\mbox{AIB~3.3\,$\upmu$m}.
These two H$_2$ lines  originate from high energy levels populated through radiative and collisional de-excitation of FUV-pumped H$_2$
\citep{Habart23a,Peeters24}.
 In this study, we rotated all images by 37.5$^o$ clockwise  to bring the FUV illumination from the Trapezium stars in the horizontal \mbox{direction}.\footnote{We adopted the convention that FUV radiation from the Trapezium stars impinges from the right-hand side of the images. This convention applies to all rotated images and PDR models  presented in this paper.}
The vertical dotted line marks the approximate position of the main H$_2$ DF, the H$_2$/H transition zone of the PDR. 
To the left of this  dotted line, the PDR gas is mostly molecular, meaning \mbox{[H$_2$]\,$\gg$\,[H]}. To the right,  the PDR gas is predominantly atomic, with
\mbox{[H$_2$]\,$\ll$\,[H]}. This  strongly irradiated atomic  zone hosts the brightest 
AIB emission \citep[][]{Habart24,Peeters24,Chown24},
an indication of how resistant these aromatic species can be.
 The  DF runs roughly parallel to the ionization front. 
The rim of the AIB emission delineates the IF, the edge of the \HII~region.
 
Instead of a unique H$_2$/H transition zone,  Figs.~\ref{fig:cch_c18o_jwst_integrated_intensity}c and \ref{fig:cch_c18o_jwst_integrated_intensity}e 
show a very corrugated  zone 
composed of multiple small-scale DFs \citep[][]{Habart23a,Habart24,Peeters24}. These IR \mbox{H$_2$-emitting} fronts nearly match the \mbox{HCO$^+$ $J$\,=\,4--3} structures
previously observed by ALMA \citep[][]{Goico16}. On the other hand, the more FUV-shielded cloud interior (as traced by the optically thin \mbox{C$^{18}$O} emission) shows a less filamentary but clumpier morphology, which becomes bright as the IR H$_2$  emission dims. This implies that the H$_2$ emission from DF3 and DF4 marks the FUV-irradiated rims of  molecular gas structures characterized by bright \mbox{C$^{18}$O} emission
(see also \mbox{Fig.~\ref{fig:contours}}).

Our  ALMA images reveal  bright [\CI]\,609\,$\upmu$m emission  very close  to many H$_2$ emission peaks (Fig.~\ref{fig:cch_c18o_jwst_integrated_intensity}d). 
In addition, a more diffuse [\CI]\,609\,$\upmu$m emission component exists toward the  cloud interior.
The ALMA images also reveal bright,
filamentary C$_2$H emission in all small-scale DFs, 
either very close to or nearly coincident with the H$_2$ emission (Fig.~\ref{fig:cch_c18o_jwst_integrated_intensity}a). 
Indeed, the spatial distribution of the C$_2$H emission is well
correlated with that of \mbox{FUV-pumped H$_2$} (traced by the excited
\mbox{$v$\,=\,1--0 $S(1)$} and \mbox{$v$\,=\,0--0 $S$(9)} lines; \mbox{see Appendix~\ref{sec:correlations}}).
In addition, $I$(C$_2$H\,4--3) is anti-correlated with the 3.3\,$\upmu$m
AIB emission in the atomic PDR zone, where the AIB emission is much brighter 
(\mbox{Appendix~\ref{sec:correlations}}).

The molecular emission  in the  atomic PDR  (to the right of the vertical
dotted line in Fig.~\ref{fig:cch_c18o_jwst_integrated_intensity}) is more difficult to interpret. It can arise from the Bar, but also from the background, from 
deeper layers of  \mbox{OMC-1}  illuminated at a slanted angle.
Determining its origin requires an investigation of the line velocity centroid, which may vary among these components (\mbox{Sect.~\ref{sec:channel maps}}).

\begin{figure}[t]
\centering   
\includegraphics[scale=0.30, angle=0]{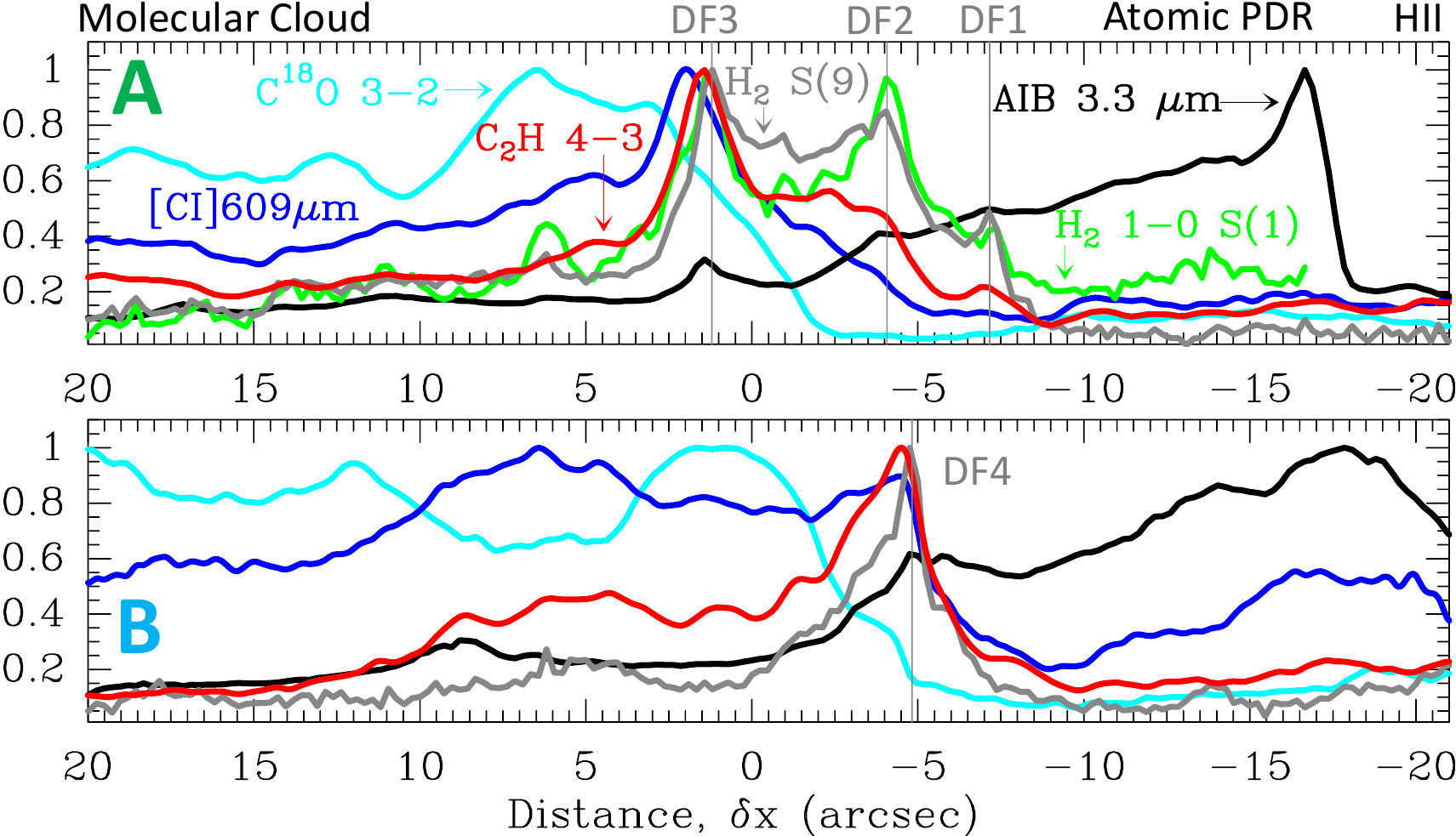}
\caption{Vertically averaged crosscuts A and B 
perpendicular to the Bar and parallel to the FUV illumination direction 
(Fig.~\ref{fig:cch_c18o_jwst_integrated_intensity}c). 
Crosscut~A (B) passes through \mbox{$\delta y$\,=\,$+$2$''$} (\mbox{$\delta y$\,=\,$-$20$''$}). Both crosscuts have a width of \mbox{$\Delta(\delta y)$\,=\,6$''$. Both plots show normalized line intensities.}}
\label{fig:crosscuts}
\end{figure}
\begin{figure*}[t]
\centering   
\includegraphics[scale=0.67, angle=0]{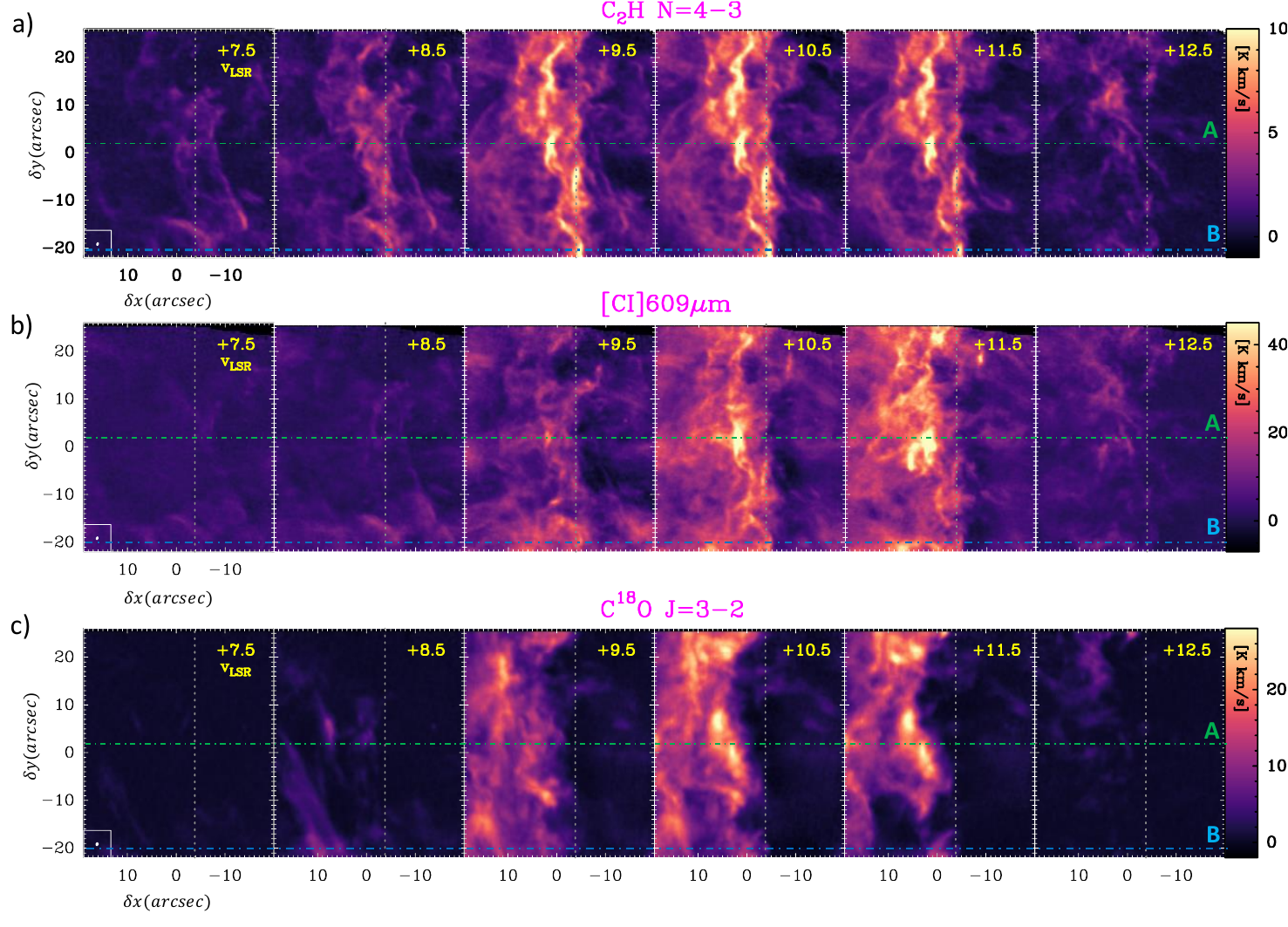}
\caption{ALMA velocity channel maps of the Bar from 
$v_{\rm LSR}$\,$=$\,$+$7 to $+$12.5\,km\,s$^{-1}$
in bins of 1\,km\,s$^{-1}$.
All images have been rotated to bring the FUV-illuminating direction into the
horizontal direction (from the right).
The synthesized beam of each mosaic is indicated in the bottom-left corner of the first panel.
The vertical dashed white line crosses DF2 (\mbox{$\delta x$\,$\simeq$\,$-$4$''$}) and marks  the approximate position of the main H$_2$ DFs parallel to the Bar.
The horizontal lines  show the position of the vertically averaged cuts A and B,
parallel to the incoming FUV,
discussed in the text.}
\label{fig:collage_vel_chan}
\end{figure*}

\subsection{Crosscuts A and B through the photodissociation region}
\label{sec:cuts_A_B}

To dissect the typical structures and  stratification seen in the PDR, \mbox{Fig.~\ref{fig:crosscuts}} shows normalized intensity cuts perpendicular to the  Bar,  extracted from  ALMA and IR filter images 
(Fig.~\ref{fig:cch_c18o_jwst_integrated_intensity}c indicates the  position
of these cuts).
We chose two crosscuts passing through  $\delta y$\,=\,$+$2$''$ (cut~A) and $\delta y$\,=\,$-$20$''$ (cut~B), roughly parallel to the incoming FUV radiation. 
The sharp drop in the 3.3\,$\upmu$m AIB emission 
delineates the location of the IF \citep[see][]{Peeters24}.
In cut~A, three differentiated \mbox{(H$_2$-bright)} DFs appear at
 \mbox{$\sim$\,11$''$ (DF1)}, \mbox{$\sim$\,14$''$ (DF2)}, and  \mbox{$\sim$\,17$''$ (DF3)} from the IF \citep[at $\sim$\,0.021\,pc, $\sim$\,0.027\,pc, and $\sim$\,0.033\,pc, respectively; see also,][]{Habart24,Peeters24}.  
 DF1 and DF2 show faint [\CI] and C$_2$H emission relative to that of H$_2$, and they do not show significant C$^{18}$O emission.  Thus, they appear as filaments that are translucent to FUV radiation. That is,  they show H$_2$ and AIB 3.3 $\mu$m emission but 
 have low column density perpendicular to the line of sight (no or very little C$^{18}$O). 
In \mbox{Sect.~\ref{subsec:geometry}}, we infer  more details about their geometry and possible origin.

However, DF3 is different. It is located deeper in the molecular PDR and represents a DF type corresponding to the irradiated rim of a larger molecular  structure or clump, which is traced by bright C$^{18}$O emission.
The separation between the [\CI]\,609\,$\upmu$m and H$_2$ emission peaks
(roughly the separation between the CO/C and H$_2$/H transition zones)
is very small $\lesssim$\,0.8$''$  \mbox{($\lesssim$\,0.002\,pc\,$\simeq$\,400\,au)} and
implies  \mbox{$\Delta A_V$\,$\lesssim$\,0.1--0.2\,mag}, adopting  \mbox{$n_{\rm H}$\,$\simeq$\,(5--10)$\times$10$^{4}$\,cm$^{-3}$}
\citep[as in the atomic PDR; ][and references therein]{Peeters24} and  \mbox{$A_V$/$N_{\rm H}$\,=3.5$\times$10$^{-22}$ mag\,cm$^2$}  \citep{Cardelli89}. This  width is smaller than the expected separation in a constant-density PDR  \citep[\mbox{$\Delta A_V$\,$\simeq$\,1--2\,mag};][]{Tielens85}, and 
suggests higher densities in the molecular PDR compared to the atomic PDR zone.

\begin{figure*}[ht]
\centering   
\includegraphics[scale=0.865, angle=0]{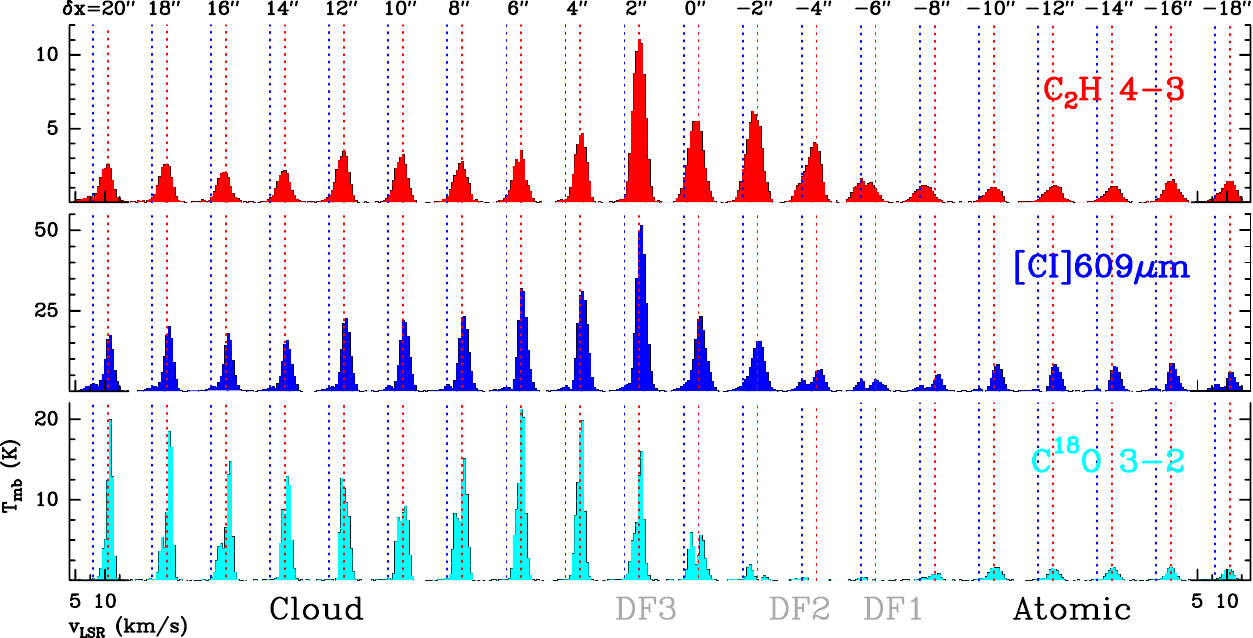} 
\caption{Spectra along cut A in the direction of FUV illumination from the Trapezium 
(from the right). The figure shows spectra averaged over 2$''$$\times$2$''$ boxes.
The red dashed lines mark typical velocity centroid of the Bar, 
at $v_{\rm LSR}$\,$\simeq$\,10.5\,km\,s$^{-1}$. The blue dashed lines mark the typical velocity centroid of the background \mbox{OMC-1} emission
at $v_{\rm LSR}$\,$\simeq$\,8\,km\,s$^{-1}$. This is a minor component of the total line intensity within the PDR. } 
\label{fig:spectra_cutA}
\end{figure*}

Cut~B shows a single H$_2$-bright DF (DF4) that 
nearly coincides with the bright C$_2$H emission. This cut shows an even less pronounced spatial stratification.
This suggests that DF4 is the rim of a high density structure,
 more akin to a tilted sheet of
\mbox{FUV-irradiated}  gas. Deeper into the  cloud, cut~B reveals moderately bright [\CI]\,609\,$\upmu$m emission (i.e., it does not disappear) that follows that of \mbox{C$^{18}$O~$J$\,=\,3--2} at \mbox{$\delta x$\,$>$\,10$''$}. 
This deeper [\CI]\,609\,$\upmu$m emission seems to be associated with a secondary peak at 
$\delta x$\,$\simeq$\,$+$7$''$. 
A relevant result for hydrocarbon chemistry is that in all DFs, \mbox{C$_2$H} peaks
ahead of the [\CI]\,609\,$\upmu$m emission peak, implying that 
the observed C$_2$H emission arises near the \mbox{CO\,/\,C\,/\,C$^+$} transition, where the gas is rich in C$^+$ ions.

\subsection{Small-scale gas kinematics}
\label{sec:channel maps}

Figure~\ref{fig:collage_vel_chan} shows   C$_2$H $N$\,=\,4--3 (a), [\CI]\,609\,$\upmu$m (b), and 
C$^{18}$O $J$\,=\,3--2 (c) emission in different LSR velocity intervals. These plots dissect the line emission in 1~km\,s$^{-1}$ channels, from $v_{\rm LSR}$\,=\,$+$7.5 to $+$12.5~km\,s$^{-1}$.  
They unveil the small-scale structure, both spatially and in velocity, of the molecular gas  exposed to strong FUV radiation. 
They show many  
C$_2$H--emitting elongated structures roughly paralel to the IF.
The morphological similitude between the H$_2$ emission
and the edge of the C$_2$H emission suggests that C$_2$H is a good proxy  of the \mbox{H$_2$-emitting} gas kinematics (see also the next section and \mbox{Fig.~\ref{fig:PDR_correlations}}).

Emission from the Bar PDR typically peaks at LSR velocities around
\mbox{$v_{\rm LSR}$\,$\simeq$10--11\,km\,s$^{-1}$} \citep[e.g.,][]{Cuadrado17}. The  LSR velocity of the \mbox{OMC-1} emission, in the background,
is \mbox{$v_{\rm LSR}$\,$\simeq$\,8--9\,km\,s$^{-1}$}
\citep[e.g.,][]{Berne14,Goico20}. 
\mbox{Figure~\ref{fig:spectra_cutA}} shows velocity-resolved line profiles extracted across \mbox{cut A}. The integrated line intensities in DF3 and deeper molecular layers are largely dominated by emission from the Bar (red vertical dashed lines). 
Some spectra show a minor contribution from a second, faint emission component at \mbox{$v_{\rm LSR}$\,$\simeq$\,8\,km\,s$^{-1}$} (blue vertical dashed lines). This faint component becomes more significant in DF1 (at \mbox{$\delta x$\,$\simeq$\,$-$7$''$}) and it originates from gas in \mbox{OMC-1}, in the background, or from gas structures behind the PDR, likely the base of the cloud escarpment that forms the Bar. 
A two-Gaussian fit to the [\CI]\,609\,$\upmu$m spectra (where the two components are more clearly seen) reveals that the velocity centroid of the
\mbox{$v_{\rm LSR}$\,$\simeq$10--11\,km\,s$^{-1}$} component (the Bar PDR) remains approximately constant throughout the Bar
(\mbox{Fig.~\ref{fig:cut_A_centroids}}). 
 C$^{18}$O  shows minimal contribution from the \mbox{$v_{\rm LSR}$\,$\simeq$\,8\,km\,s$^{-1}$} component \mbox{(OMC-1)}, but most positions exhibit a faint  component at $\sim$9.5\,km\,s$^{-1}$. This suggests either the presence of two sub-components in the Bar or double-peaked profiles caused by slow radial motions.
Finally, we detect  emission within the atomic PDR zone  at the LSR velocity of the Bar. This emission is more difficult to interpret as molecular abundances in this zone are expected to be \mbox{negligibly low (\mbox{Sect.~\ref{sec:pdr_mods}})}.

\section{Analysis: Zoom-in of the dissociation fronts}
\label{sec:analysis}

\begin{figure}[h]
\centering   
\includegraphics[scale=0.445, angle=0]{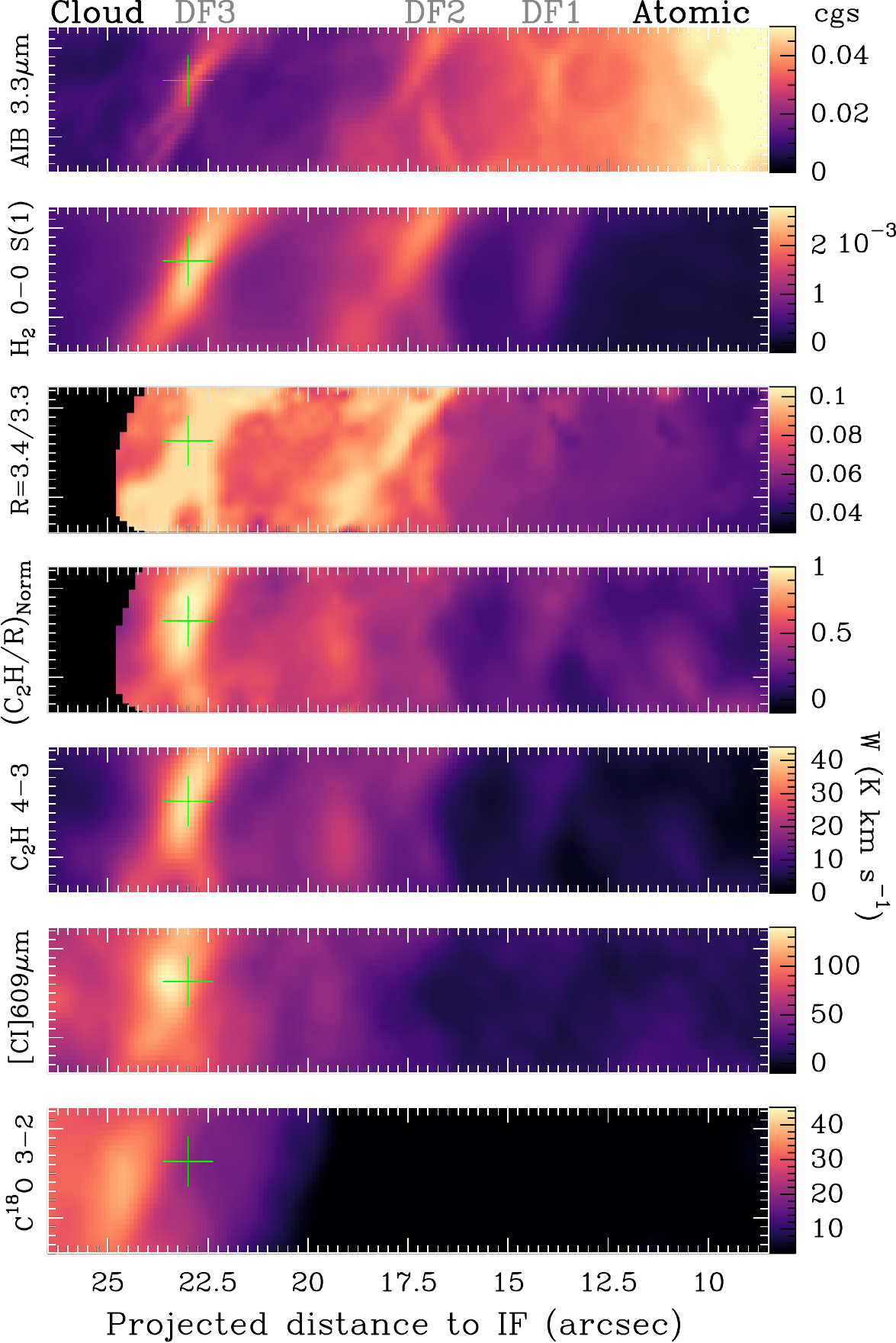}
\caption{Small FoV 
observed with NIRSpec and MIRI-MRS.
The green cross shows
a \mbox{C$_2$H~$N$\,=\,4--3} emission peak in DF3 region.
This peak nearly coincides with the IR H$_2$ emission peak.
``cgs'' refers to \text{erg\,s$^{-1}$\,cm$^{-2}$\,sr$^{-1}$}.
The \mbox{(C$_2$H/R)$_{\rm Norm}$} map refers to the normalized
 \mbox{C$_2$H~$N$\,=\,4--3} line intensity divided by the normalized \mbox{3.4/3.3\,$\upmu$m AIB} ratio.
 Figure~\ref{fig:PDR_MIRI_crosscut} shows vertically averaged intensity 
cuts of this field.}
\label{fig:MIRI_images}
\end{figure}
 
Figure~\ref{fig:MIRI_images} shows a zoom-in of a small FoV ($\sim$18$''$\,$\times$\,4$''$) 
observed with NIRSpec and MIRI-MRS. \mbox{Mid-IR}  observations probe the lower-energy H$_2$ pure-rotational lines. 
To first order, the intensity of these collisionally excited \mbox{H$_2$ $v$\,=\,0--0}   \mbox{low-$J$}  lines  (\mbox{Fig.~\ref{fig:MIRI_images}}) is proportional
to the column density of warm molecular hydrogen, $N_{\rm warm}$(H$_2$) 
\citep[\mbox{Sect.~\ref{sec:cut_C_H2}} and][]{Putte24}. The  green cross in Fig.~\ref{fig:MIRI_images} marks the \mbox{C$_2$H~$N$\,=\,4--3} emission peak in DF3. This peak nearly coincides with the IR H$_2$  emission peak.

\subsection{$N$(H$_2$)$_{\rm warm}$ and $T_{\rm rot}$(H$_2$) across the Bar\label{sec:cut_C_H2}}

\begin{figure*}[t]
\centering   
\includegraphics[scale=0.455, angle=0]{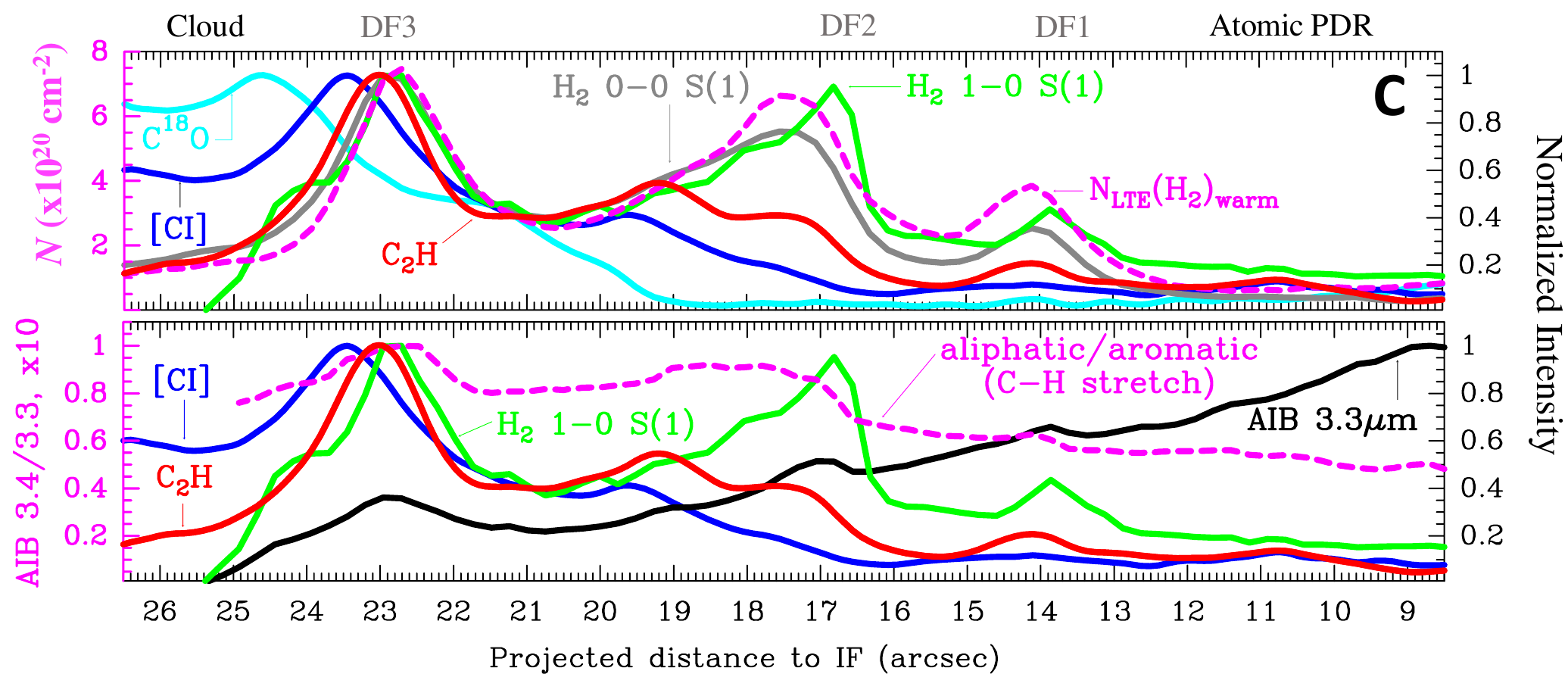} 
\caption{Vertically averaged crosscut C with \mbox{$\Delta(\delta y)$\,=\,2$''$}. 
This cut passes through the green cross  in \mbox{Fig~\ref{fig:MIRI_images}}.
In the upper panel, the magenta dashed curve shows 
the  column density of warm H$_2$ obtained from \mbox{$v$=0--0 $S$(4)} and  $S$(2) 
lines, observed with MIRI-MRS, assuming a Boltzmann distribution at  $T_{64}$ ($\simeq$\,600\,K, shown in Fig.~\ref{fig:cut_c_T42}).
In the bottom panel, the magenta dashed curve shows the \mbox{3.4\,/\,3.3\,$\upmu$m}
AIB intensity ratio ($\times$10).}
\label{fig:PDR_MIRI_crosscut}
\end{figure*}

Figure~\ref{fig:PDR_MIRI_crosscut} displays the vertically averaged intensity crosscut~C
extracted  along the small FoV   observed with JWST spectrometers (Figs.~\ref{fig:cch_c18o_jwst_integrated_intensity}c and  ~\ref{fig:MIRI_images}). 
Pure rotational H$_2$ $v$\,=\,0--0 $S(2)$ and $S(4)$ lines at 12.3\,$\upmu$m and
8.1\,$\upmu$m are the lowest energy \mbox{\textit{para}-H$_2$} lines observed by \mbox{MIRI-MRS}.
The intensities of these lines are much less affected by foreground extinction 
 than the
\mbox{$ortho$-H$_2$ $S(3)$} line at 9.6\,$\upmu$m (i.e., within the 9.7\,$\upmu$m silicate grain absorption feature). 
 Assuming optically thin H$_2$ line emission and no foreground extinction, one can convert the observed 
H$_2$ \mbox{$v$\,=\,0--0 $S$(4)} and $S$(2)  line intensities into
$J$\,=\,6 and $J$\,=\,4 level column densities ($N_6$ and $N_4$),  and derive the rotational
temperature $T_{\rm 64}$ as
\begin{equation}
T_{\rm 64}\,=\,\frac{-\Delta E_{\rm 64}/k}{{\rm ln}\,(N_6\,g_4\,/\,N_4\,g_6)},
\end{equation}
where $\Delta E_{\rm 64}/k$ is the energy level difference, and $g$ are the level degeneracies. This calculation leads to  peak $T_{\rm 64}$ values of $\sim$600\,K in DF2 and DF3
\citep[see also][]{Putte24}.
 \mbox{Figure~\ref{fig:cut_c_T42}} shows the resulting $T_{\rm 64}$  profile across cut~C. 
 Interestingly, \mbox{$I$(C$_2$H\,4--3)} does not follow the $T_{\rm 64}$ profile 
(see \mbox{Fig.~\ref{fig:cut_c_T42}}). Instead,  
 it more closely follows the H$_2$ emission and  is well correlated with the FUV-pumped H$_2$ emission (see \mbox{Appendix~\ref{sec:correlations}}).

 The $S$(2) and $S$(4)  lines have relatively low critical densities ($n_{\rm cr}$) for H$_2$--H$_2$ inelastic collisions 
\citep[from a few 10$^3$ to several 10$^5$\,cm$^{-3}$ at $T$\,$\simeq$\,600\,K;][]{Wan18,Hernandez21}. This $n_{\rm cr}$ is comparable or lower than the gas density in the DFs \citep{Peeters24}. Thus, $T_{\rm 64}$ is a good proxy of the gas temperature in the \mbox{IR H$_2$--emitting} gas  (only if 
\mbox{$n_{\rm H_2}$\,$\ll$\,$n_{\rm cr}$}, then 
\mbox{$T_{\rm 64}$\,$\ll$\,$T_{\rm k}$}).
We also estimated
the column density of warm H$_2$  along each line of sight in crosscut~C.
Assuming a \mbox{Boltzmann} population of the $N_4$ and $N_6$
levels at $T_{\rm 64}$, we determine
\mbox{$N_{\rm warm}$($p$-H$_2$)$_{\rm LTE}$} as
\begin{equation}
N_{\rm warm}(p-{\rm H_2})_{\rm LTE}\,=\,\frac{N_4}{g_4}\,Q(T_{\rm 64})\,e^{+E_4/kT_{\rm 64},} 
\end{equation}
where $Q(T)$ is the  rotational partition function of $p$-H$_2$.
The total column density of warm H$_2$, \mbox{$N_{\rm warm}$(H$_2$)$_{\rm LTE}$}, is equal to
\mbox{$N_{\rm warm}$($p$-H$_2$)$_{\rm LTE}$\,(1\,+\,OTP)}, where OTP is the H$_2$ 
ortho-to-para ratio. The H$_2$ OTP ratio across the Bar  is fairly constant and equal to three
\citep[][Sidhu in prep.]{Putte24}. 

Figure~\ref{fig:PDR_MIRI_crosscut} shows the resulting
\mbox{$N_{\rm warm}$(H$_2$)$_{\rm LTE}$} profile (dashed magenta curve)
across cut~C. This curve
peaks at DF3, DF2, and DF1, roughly following the low-energy \mbox{H$_2$ $v$=0--0 $S$(1)} emission profile.  
In reality, foreground and internal extinction in the PDR dim the IR H$_2$ line emission.
Although this attenuation is not large toward Orion, in particular toward the edge of the
cloud, we  included a line intensity extinction correction as
$I_{\rm corr}\,=\,I_{\rm obs}\,e^{+\tau_{\lambda}}$, with $\tau_{\lambda}\,=\,A_{\lambda}/1.086$
and $A_{\lambda}$/$A_{V}$\,=\,0.045  at the wavelength of the $S(2)$ and $S(4)$ lines  \citep[appropriate to Orion;][]{Decleir22,Gordon23}.  
Table~3 provides the corrected \mbox{$N$(H$_2$)$_{\rm warm}$} values
toward DF3, DF2, and DF1 using two  extinction estimations: those
of \cite{Peeters24} and  \cite{Putte24}. These calculations yield a reasonable range
of total column densities in these DFs, assuming \mbox{$N$$_{\rm H}$\,$\simeq$\,$N$(H)\,+\,2\,$N$(H$_2$)$_{\rm warm}$} and
 \mbox{$N$(H)\,$\simeq$\,$N$(H$_2$)$_{\rm warm}$}. We obtain
  $N$$_{\rm H}$ values of \mbox{2.61--4.73}, \mbox{2.33--3.33}, and \mbox{4.92--2.75} times 10$^{21}$\,cm$^{-2}$ in DF3, DF2, and DF1, respectively for the two
extinction corrections mentioned above.

\subsection{Geometry and origin of the dissociation fronts}
\label{subsec:geometry}

Adopting a typical gas density, $n_{\rm H}$, of a few \mbox{10$^{5}$\,cm$^{-3}$} 
in the DFs, based on an analysis of the mid-IR H$_2$ emission \citep{Putte24}
and consistent with \mbox{Sect.~\ref{sec:mtc_mods}}--which assumes that the 
\mbox{C$_2$H 4--3} and H$_2$ rotational emissions are nearly 
co-spatial\footnote{As observed
(e.g., Fig.~\ref{fig:PDR_MIRI_crosscut}) and predicted by our PDR model 
(Fig.~\ref{fig:PDR_line_emissivity}).}--the inferred $N_{\rm H}$ columns imply a  scale size along the line of sight \mbox{(l-o-s)} of \mbox{$l_{\rm los}$\,=\,$N_{\rm H}$\,/\,$n_{\rm H}$\,$\simeq$\,10$^{16}$\,cm} (\mbox{$\sim$0.003\,pc}).
As DF1 and DF2 show either no or  very faint C$^{18}$O emission, they are translucent to FUV radiation, indicating extinction depths in the plane of the sky ($\delta x$; the FUV illumination direction) of  \mbox{$\Delta$$A_V$\,$\lesssim$\,1 mag}, or  \mbox{$N_{\rm H,\,\delta x}$} of a few \mbox{10$^{21}$\,cm$^{-2}$}, which gives 
\mbox{$l_{\rm \delta x}$\,$\lesssim$\,10$^{16}$\,cm}. These $\delta x$ and l-o-s spatial scale sizes
\mbox{($\sim$1$''$--2$''$, in agreement with the observed emission width in $\delta x$)} are smaller than the size of the elongated H$_2$--emitting structure in  the  $\delta y$ direction. 
Therefore, one possibility is that DF1 and DF2 are true small-scale filaments, whose lengths are significantly greater than their widths.
Since they run roughly parallel to the IF, they may represent the effects of a shockwave propagating into the molecular cloud, driven by  FUV radiation  \citep[e.g.,][]{Hill78,Bron18} and perhaps  by stellar winds from \mbox{$\theta^1$\,Ori C} \citep{Pabst19,Pabst20}. This shockwave would lead  to localized gas compression and  minor density perturbations \citep[see also][]{Goico16}.
Another possibility is that these DFs represent a terraced-field-like cloud structure 
\citep[i.e., multiple cloud surfaces;][]{Habart24,Peeters24}, with several steps seen from above, to account for the succession of nearly edge-on DFs. The lack of significant C$^{18}$O emission in DF1 and DF2, however, would imply that gas densities are much lower than in DF3 and DF4.
Hydrodynamical simulations
will be needed to confirm these two scenarios (filaments and sheets vs. cloud surfaces) as well as their origin.

DF3 and other \mbox{H$_2$-emitting} structures that border the molecular cloud traced by bright C$^{18}$O\,3--2 emission (\mbox{Fig.~\ref{fig:contours}}) correspond to FUV-illuminated rims of larger, likely denser molecular gas structures or clumps \cite[][]{Lis03}. In DF3, \mbox{$l_{\rm los}$(warm H$_2$)} is also $\sim$10$^{16}$\,cm, which is significantly smaller than the projected size of the C$^{18}$O-emitting structure
(several arcseconds). 
This suggests that the IR H$_2$ emission traces only the limb of roughly spherical structures, which may be remnants of cloud turbulence
\citep[e.g.,][]{Hartman07,Glover11,Federrath13} or produced by photoevaporation processes \citep[e.g.,][]{Gorti02}.

\begin{table*}[h] \label{table:columns-H}
  \begin{center}
    \caption{Warm H$_2$ and total column densities in selected  DF
    positions of the Orion Bar PDR.}
     \begin{tabular}{c   c c c c c c @{\vrule height 8pt depth 5pt width 0pt}}   
      \hline \hline
             & $N$(H$_2$)$_{\rm warm}$  & $N$$_{\rm H}^{(a)}$ &  $N$(H$_2$)$_{\rm warm}$ & $N$$_{\rm H}^{(a)}$ & $N$(H$_2$)$_{\rm warm}$ & $N$$_{\rm H}^{(a)}$ \\   
  Position   &  [cm$^{-2}$]             & [cm$^{-2}$]   &  [cm$^{-2}$]             & [cm$^{-2}$]   & [cm$^{-2}$]  & [cm$^{-2}$]  \\ \hline
& \multicolumn{2}{c}{\underline{No extinction correction}} & \multicolumn{2}{c}{\underline{Correction 1$^{(b)}$}} & \multicolumn{2}{c}{\underline{Correction 2$^{(c)}$}} \\
    DF1      &   3.50$\times$10$^{20}$   & 1.05$\times$10$^{21}$  & 1.64$\times$10$^{21}$  & 4.92$\times$10$^{21}$ & 9.07$\times$10$^{20}$ & 2.72$\times$10$^{21}$ \\
    DF2      &   5.42$\times$10$^{20}$   & 1.63$\times$10$^{21}$  & 7.76$\times$10$^{20}$  & 2.33$\times$10$^{21}$ & 1.11$\times$10$^{21}$ & 3.33$\times$10$^{21}$ \\
    DF3      &   7.47$\times$10$^{20}$   & 2.24$\times$10$^{21}$  & 8.71$\times$10$^{20}$  & 2.61$\times$10$^{21}$ & 1.58$\times$10$^{21}$ & 4.73$\times$10$^{21}$\\ 
\hline
\end{tabular}
\tablefoot{$^{(a)}$Total column density with respect to H nuclei assuming \mbox{$N$$_{\rm H}$\,$\simeq$\,$N$(H)\,+\,2$N$(H$_2$)$_{\rm warm}$} and
 \mbox{$N$(H)\,$\simeq$\,$N$(H$_2$)$_{\rm warm}$} in these DFs.
$^{(b)}$Extinction determined in the intermingled formalism by \citet{Peeters24}. The total
(foreground and internal) visual extinction toward
 DF1, DF2 and DF3 positions is 37.3, 8.64, and 3.70\,mag, respectively (see their \mbox{Table~1}).
$^{(b)}$Extinction determined by \citet{Putte24} assuming a screen geometry and 23, 14, and 18\,mag 
of visual extinction toward DF1, DF2, and DF3,
respectively.}
\end{center}
\end{table*} 

\subsection{C$_2$H  emission and the 3.4\,/\,3.3\,$\upmu$m  band ratio}
\label{sec:cut_C_carbon}

The C$_2$H emission approximately follows local enhancements of the  
3.3\,$\upmu$m AIB intensity, which is proportional to the abundance of the carriers, to the total column density $N_{\rm H}$, and  to the local flux of FUV photons \citep[e.g.,][]{Habart24}.
Interestingly, the spatial distribution of the C$_2$H emission more closely resembles the observed peaks of the 3.4/3.3\,$\upmu$m AIB intensity  ratio 
(\mbox{Figs.~\ref{fig:MIRI_images} and  \ref{fig:PDR_MIRI_crosscut}}).
The 3.4\,$\upmu$m AIB emission is generally  assigned to C--H stretching mode of a small quantity of H atoms
bonded to $sp$$^3$ C atoms, either in hydrocarbon radical groups 
\citep[mostly \mbox{--CH$_3$}, so-called methylated-PAHs e.g.,][]{Jourdain90,Joblin96}
 or super-hydrogenated PAHs \citep{Schutte93,Bernstein96}. The latter are not really expected to dominate in the Bar, as any extra H atom will be quickly photo-detached \citep[when \mbox{$G_0$/$n$(H)\,$>$\,0.03};][]{Andrews16}.  
Other studies associate the   \mbox{3.4\,/\,3.3\,$\upmu$m} AIB ratio
with the hydrogenation levels of carbonaceous ``nanograins'' 
\mbox{\citep[][]{Elyajouri24}}.
Regardless of the nomenclature, the carriers are in the molecular domain and must be highly excited to emit at 3.4\,$\upmu$m. 
Here, we attribute the  \mbox{3.4\,/\,3.3\,$\upmu$m} AIB ratio 
to the aliphatic-to-aromatic content of PAHs  \citep[e.g.,][]{Joblin96,Pilleri15,Li12,Yang16,Peeters24,Schroetter24}. 
This is also supported by the presence of weak bands at $\sim$6.9\,$\upmu$m in DF3 and DF2, likely originating from CH deformation modes of aliphatic groups \citep[][]{Chown24}.

The aliphatic \mbox{C--H} bonds are easier to dissociate than the aromatic \mbox{C--H} bonds in the PAH skeleton
\citep{Marciniak21}.
Indeed, the  \mbox{3.4\,/\,3.3\,$\upmu$m} AIB intensity  ratio is remarkably low 
($\sim$\,0.04) in the atomic PDR \citep{Peeters24,Chown24,Schroetter24,Pasquini24}, which is exposed to a stronger  FUV field (higher flux and photon energy) than the DFs. 
The \mbox{3.4\,/\,3.3\,$\upmu$m} AIB   ratio increases up to  $\sim$\,0.1 in DF3
\citep[still modest compared to PDRs of lower $G_0$; e.g.,][]{Joblin96,Mori14}, where it  coincides
with the maximum   value of  \mbox{$I$(C$_2$H\,4--3)}. In general, the  \mbox{3.4/3.3\,$\upmu$m} AIB  ratio  follows the C$_2$H
emission profile  (see also \mbox{Fig.~\ref{fig:corr_34_33_C2H}}), where  local peaks of  the ratio 
coincide with local peaks of \mbox{$I$(C$_2$H\,4--3)}. 
This implies that the conditions triggering the formation of simple hydrocarbons also favor the formation (and survival) of PAHs with aliphatic side groups.
Furthermore, the enhanced abundances of related small hydrocarbon radicals
 at these peaks may also indicate a causal 
relationship\footnote{\mbox{Figure~\ref{fig:MIRI_images}} shows  the normalized C$_2$H\,4--3 line intensity divided by the normalized  \mbox{3.4/3.3\,$\upmu$m} AIB ratio. This quantity peaks in DF3. The \mbox{3.4/3.3\,$\upmu$m}  ratio globally decreases from the DFs to the atomic PDR 
\citep[due to the photoerosion of aliphatic side groups; ][]{Peeters24,Chown24}, but it declines less steeply than \mbox{$I$(C$_2$H 4--3)}. Thus, this quantity  \mbox{remains $<$\,1}
and decreases from DF3 to the atomic PDR.} 
between these radicals and the  aliphatic content of PAHs (see \mbox{Sect.~\ref{sub:link_with_aliphatic}}).

\begin{figure*}[t]
\centering    
\includegraphics[scale=0.46, angle=0]{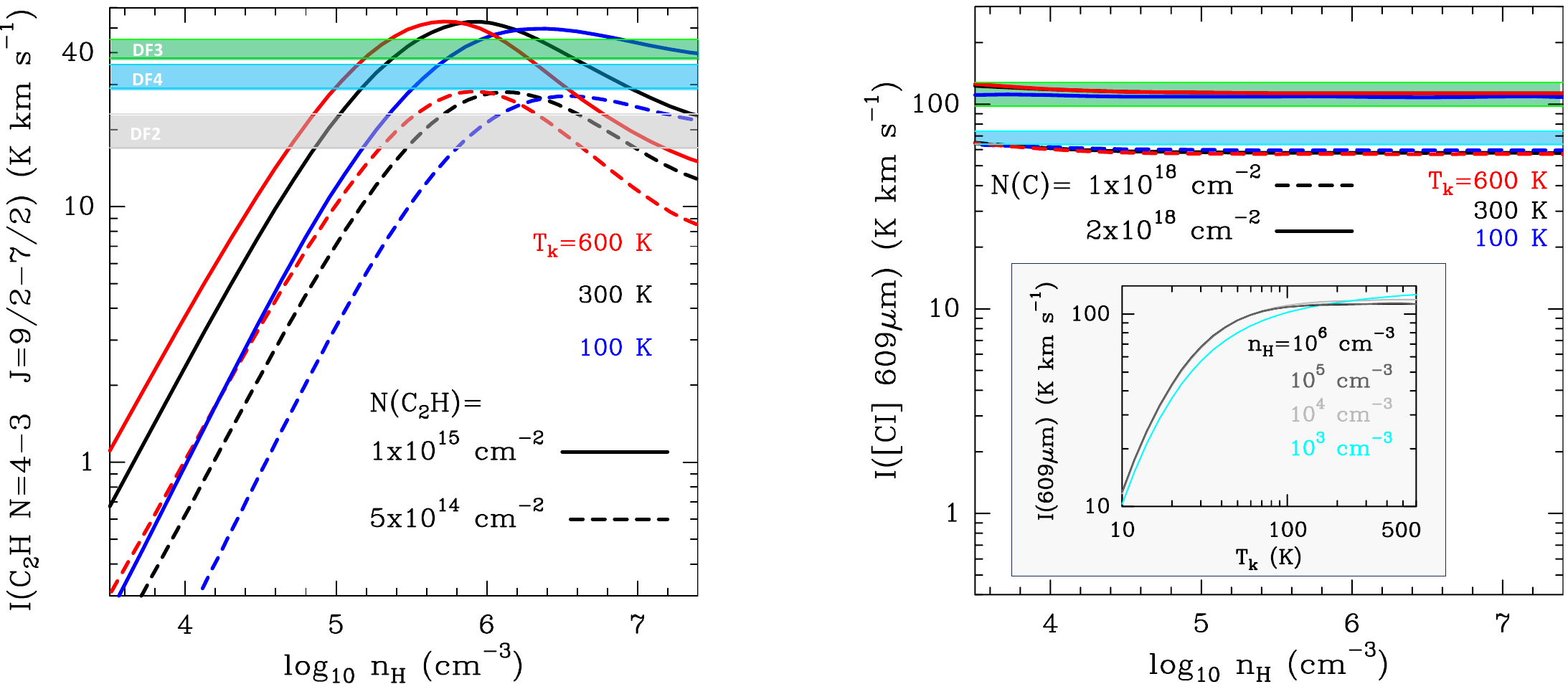}
\caption{Results from  nonlocal and non-LTE radiative transfer models
of C$_2$H (left) and [\CI] (right) for different values of $T_{\rm k}$ and $n_{\rm H}$ (single-component models$^6$).
 The horizontal green, blue, and gray shaded areas 
 show the observed peak intensities and  1$\sigma$ dispersions.}
\label{fig:MTC_model}
\end{figure*}

\subsection{C$_2$H abundance, $n_{\rm H}$, and $T$ at the radical peaks } 
\label{sec:mtc_mods}

Here we investigate the range of gas density, temperature, and C$_2$H column density 
that reproduces the peak \mbox{$I$(C$_2$H $N$~\,=\,4--3)} line intensities 
in the main DFs.
To do so, we updated our \mbox{nonlocal} and non-LTE
radiative transfer Monte Carlo model \citep[][]{Goico22} 
to treat the rotational excitation of C$_2$H  by inelastic collisions 
with \mbox{$o$-H$_2$($J$\,=\,1)}, \mbox{$p$-H$_2$($J$\,=\,0)}, and e$^-$ 
(for details, see \mbox{Appendix~\ref{App:C2H_collisonal}}). We set the  molecular gas fraction, \mbox{$f$(H$_2$)\,=\,2$n$(H$_2$)/$n_{\rm H}$}, to 2/3,
with \mbox{$n_{\rm H}$\,=\,$n$(H)\,+\,2$n$(H$_2$)}. This choice implies \mbox{$n$(H$_2$)\,=\,$n$(H)}, which is appropriate for a DF.
We also include  radiative excitation by  dust continuum photons \mbox{(Appendix~\ref{App:C2H_collisonal})} consistent with the far-IR 
  and submillimeter continuum detected in the  Bar 
\citep[][]{Arab12,Salgado16}. 
 \mbox{Figure~\ref{fig:MTC_model}} shows a grid of 
single-component models\footnote{We run single-component models
(single  $T_{\rm k}$ and $n_{\rm H}$), but the model discretizes the emission zone into multiple slabs to account for the nonlocal character of molecular excitation and line photon transport. We include thermal, opacity, and turbulent line broadening, with  
$\sigma_{\rm turb}$\,=\,1\,km\,s$^{-1}$ (matching the observed  line-widths).
These models predict  
\mbox{$T_{\rm rot}$(C$_2$H 4–3)\,=\,20--24 K}, which is much lower
than $T_{\rm k}$ due to \mbox{non-LTE} subthermal excitation, in agreement with the excitation temperature inferred from a rotational diagram 
(\mbox{Fig.~\ref{fig:DRs_corr_c2h}}).}   where we plot 
\mbox{$I$(C$_2$H 4--3 $J$\,=\,9/2--7/2)}
as a function of $n_{\rm H}$ for different gas temperatures.
Taking \mbox{$T_{\rm k}$\,=\,600\,K}  as an  upper  limit,  
the observed \mbox{$I$(C$_2$H 4--3)} line intensities  in DF3 and DF4 can be reproduced with \mbox{$N$(C$_2$H)\,$\lesssim$\,10$^{15}$\,cm$^{-2}$} and 
$n_{\rm H}$ of a few 10$^{5}$\,cm$^{-3}$. 
The best model of DF2 implies \mbox{$N$(C$_2$H)\,$\simeq$\,5$\times$10$^{14}$\,cm$^{-2}$}.
The inferred C$_2$H columns densities make C$_2$H the most ubiquitous  of all polar hydrocarbons detected by \cite{Cuadrado15} in the  Bar.  
We determine  the peak C$_2$H abundance (\mbox{[C$_2$H]\,=\,$N$(C$_2$H)/$N_{\rm H}$})  using 
the $N_{\rm H}$ column densities in \mbox{Table~1}. 
This leads to \mbox{[C$_2$H]\,$\simeq$\,(2--4)$\times$10$^{-7}$} in DF3
and \mbox{[C$_2$H]\,$\simeq$\,2$\times$10$^{-7}$} in DF2.
These values are significantly higher than the abundances previously inferred from low-resolution observations of PDRs, which dilute the emission from small spatial-scale DFs  \citep[e.g.,][]{Fuente_1996,Wiel09,Nagy15}.

\subsection{Atomic carbon at the [\CI]\,609\,$\upmu$m peaks} 
\label{sec:mtc_CI_mods}

The [\CI]\,609\,$\upmu$m line involves a forbidden transition with a very low Einstein coefficient for spontaneous emission. In the molecular PDR, inelastic collisions of C\,($^3$P) atoms with H$_2$  dominate the excitation of the [\CI] fine-structure lines. Since high-temperature, $>$100~K, \mbox{C\,($^3$P)--$o$/$p$-H$_2$} inelastic collisional rates did not exist in the literature,
we extended the scattering calculations of \cite{Klos_18,Klos_18_corr} 
to 3000\,K (see \mbox{Appendix~\ref{App:C_rates}}). The right panel of \mbox{Fig.~\ref{fig:MTC_model}} shows a grid of [\CI] models.
Because of the resulting low excitation requirement of the [\CI]\,609\,$\upmu$m line \mbox{--low} critical density ($\simeq$\,10$^3$\,cm$^{-3}$) and low level energy separation compared to $T_{\rm k}$
close to the DFs (\mbox{$\Delta E$/$k_{\rm b}$\,=\,23.6\,K\,$\ll$\,$T_{\rm k}$})-- the [\CI]\,609\,$\upmu$m emission is optically thin, collisionally excited, and nearly thermalized 
(\mbox{$T_{\rm ex}$\,$\simeq$\,$T_{\rm k}$}). In this regime ($T_{\rm k}$\,$>$\,100 K
and $n$(H$_2$)\,$>$\,10$^4$\,cm$^{-3}$), 
\mbox{$I$([\CI]\,609\,$\upmu$m)} is proportional to $N$(C) quite irrespective of the physical conditions
(inset in \mbox{Fig.~\ref{fig:MTC_model} right}). We obtain
 \mbox{$N$(C)\,$\simeq$\,2$\times$10$^{18}$\,cm$^{-2}$} 
 and  \mbox{$\lesssim$\,10$^{18}$\,cm$^{-2}$} at the $I$([\CI]) peaks,
slightly behind  DF3 and DF4, respectively. Assuming that atomic carbon becomes the major gas-phase
reservoir of carbon at the $I$([\CI]) peak (with \mbox{[C/H]\,$=$\,$1.4$\,$\times$10$^{-4}$}; \mbox{Fig.~\ref{fig:PDR_model}}), these $N$(C) values imply that
$N_{\rm H}$ increases from several 10$^{21}$\,cm$^{-2}$ at the  H$_2$ emission peak in DF3,   to \mbox{$N_{\rm H}$\,$\simeq$\,$N$(C)\,/\,[C/H]\,$\simeq$\,10$^{22}\,$cm$^{-2}$}
at the $I$([\CI]) peak, slightly behind. This increase is consistent
with [\CI] tracing slightly deeper layers of the molecular structure or clumps associated with C$^{18}$O, and suggests a density gradient.

\section{Photodissociation region models of the main dissociation fronts in the Bar} 
\label{sec:pdr_mods}

To understand the origin of the high C$_2$H abundances in the  Bar,
we model the hydrocarbon \mbox{radical} chemistry using  version 1.7 of the  Meudon PDR code \citep{LePetit06}. 
We updated the chemical network with the reaction rates we implemented to model the chemistry of CH$_{3}^{+}$ in \mbox{d203-506} \citep{Berne23}, including the newly computed
photodissociation cross section for CH$_{3}^{+}$ \citep{Mazo24}. 
Our gas-phase chemical network includes $v$-state-dependent rate constants for reaction of
FUV-pumped ro-vibrationally excited H$_2$ (hereafter H$_{2}^{*}$) with C$^+$, O, S$^+$, S, and N
\citep[e.g.,][]{Zanchet19,Veselinova2021,Goico22b}. These nonthermal reactions play a key role in initiating the gas chemistry in dense PDRs \citep[e.g.,][]{Sternberg95}.
We also include simple gas-grain exchanges for O, OH, H$_2$O, O$_2$, C, and CO. These species adsorb on dust grains as temperatures drop, are photo-desorbed by FUV photons, desorb via cosmic-ray impacts, and thermally sublimate. Only for water ice formation, we include the grain surface reactions \mbox{s-O\,+\,s-H\,$\rightarrow$\,s-OH} and \mbox{s-OH\,+\,s-H\,$\rightarrow$s-H$_2$O}
\citep[e.g.,][]{Hollenbach09,Putaud19}, where s- refers to the species in the solid.
Our model does not include PAH chemistry.

\begin{table}[b]
\caption{Main parameters used in the PDR models of the main DFs.\label{table:PDR-mods}} 
\centering
\begin{tabular}{ccc@{\vrule height 8pt depth 5pt width 0pt}}
\hline\hline
Model parameter                                 &     Value                                     &     Note        \\ 
\hline
FUV illumination, $G_0$                                     &     2$\times$10$^4$ Habing                     &              \\
$A_{\rm V}$\,(depth into the PDR)                         &     10 mag                                                   &                 \\
Thermal pressure $P_{\rm th}/k$                 &    (0.2--2)$\times$10$^8$\,cm$^{-3}$K              &  Isobaric                 \\
Density $n_{\rm H}$\,=\,$n$(H)\,+\,2$n$(H$_2$)  &  $n_{\rm H}$\,=\,$P_{\rm th}\,/\,kT_{\rm k}$  &  Gradient          \\
\hline
$R_{\rm V}$\,=\,$A_{\rm V}$/$E_{\rm B-V}$       &  5.5                                          &  Orion$^a$      \\
Cosmic Ray $\zeta_{\rm CR}$                     & 10$^{-16}$\,H$_2$\,s$^{-1}$                         &             \\
Abundance O\,/\,H                               & 2.6$\times$10$^{-4}$                          & Orion$^b$                 \\
Abundance C\,/\,H                                                       & 1.4$\times$10$^{-4}$                                    &  Orion$^b$      \\
Abundance S\,/\,H                                                           & 1.4$\times$10$^{-5}$                                       &  Orion Bar$^c$    \\
\hline                                    
\end{tabular}
\tablefoot{$^a$\citet{Cardelli89}. $^b$\cite{Sofia04}.
$^c$\cite{Goicoechea21} and \cite{Fuente24}.}
\end{table}

\begin{figure}[t]
\centering   
\includegraphics[scale=0.46, angle=0]{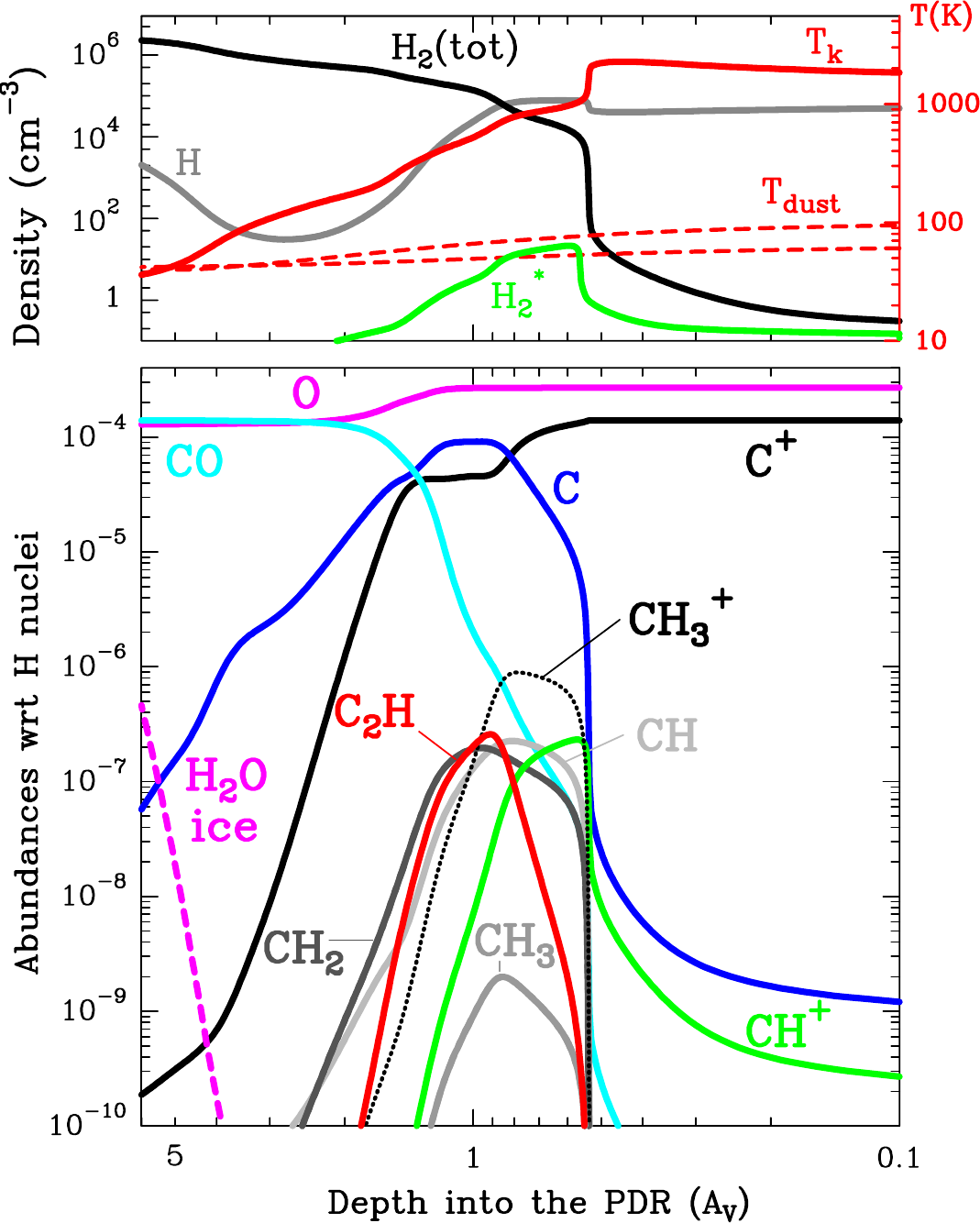}
\caption{Isobaric PDR model  with \mbox{$G_0$\,=\,2$\times$10$^4$} and \mbox{$P_{\rm th}$/$k$\,=\,10$^8$\,K\,cm$^{-3}$}. 
\mbox{\textit{Upper panel}}: Gas density, gas temperature ($T_{\rm k}$),
and grain-size distribution maximum and minimum dust temperature ($T_{\rm dust}$) profiles as a function of $A_V$ depth into the PDR. The green curve shows the density of 
\mbox{FUV-pumped} \mbox{H$_{2}^{*}$\,($v$\,$\geq$\,1)}. 
\textit{Lower panel}: Abundance profiles.}
\label{fig:PDR_model}
\end{figure}

Following our previous studies \citep[e.g.,][]{Cuadrado15,Bron18,Joblin18}, we generically model the molecular DFs as constant thermal-pressure structures (i.e., with gas density gradients). 
\mbox{Table~\ref{table:PDR-mods}} summarizes the main input parameters.
In the Bar, $G_0$   is $\sim$6$\times$10$^4$ at the IF \citep[median value of][]{Peeters24}. The exact FUV flux reaching each DF depends on
their three-dimensional structure and location with respect to the incoming FUV radiation, 
as well as on the properties of the atomic PDR.
Since the \mbox{$I$(3.3\,$\upmu$m AIB)} emission is approximately three times lower
in DF3 than in the IF (\mbox{Fig.~\ref{fig:crosscuts}}), we adopt a reference model
with a representative value of $G_0$\,=\,2$\times$10$^4$, but the main results
do not depend on the exact value.
The upper panel in \mbox{Fig.~\ref{fig:PDR_model}} shows the resulting  physical structure
for a reference model with a constant thermal pressure \mbox{$P_{\rm th}$/$k_{\rm B}$\,=\,$n_{\rm H}$\,$T_{\rm k}$\,=\,10$^8$\,K\,cm$^{-3}$},
which provides the best fit to the complete set 
of (beam-dilution corrected) \mbox{C$_2$H~$N$\,=\,1--0 to 10--9} line intensities
 (see Appendix~\ref{subsec:pdr2mtc}).

In these generic models, the low--$A_V$\,$\simeq$\,0.1\,mag  layers represent the atomic PDR, or the inter-filament environment, with \mbox{$n_{\rm H}$\,$\simeq$\,5$\times$10$^4$\,cm$^{3}$}. The $A_V$\,$\lesssim$\,1\,mag zone is more representative
of  DF2, whereas  $A_V$\,$\geq$\,1\,mag represents DF3.
The green curve shows
the density profile of \mbox{FUV-pumped}  \mbox{H$_{2}^{*}$\,($v$$\geq$1)},
which traces the steep rise in H$_2$ abundance upon entering the DF.
 The lower panel shows the abundance profiles of gas-phase CO, C, and C$^+$, along with those
of C$_2$H, CH$^+$, CH$_{3}^{+}$, and related   radicals (CH, CH$_2$, and CH$_3$).
In agreement with our C$_2$H  observations, the predicted abundance of these radicals
reaches a maximum at the DF, where \mbox{$n$(H)\,$\simeq$\,$n$(H$_2$)}.
Hence, our term 
`hydrocarbon radical peak'. The  dashed  curve shows the water-ice abundance profile.
Given the high $G_0$, dust temperatures and photodesorption rates are large enough to prevent the formation of abundant water ice in the DFs, which would otherwise deplete the volatile oxygen.
Therefore, the gas in the DFs is oxygen-rich, 
meaning \mbox{[C]/[O]\,$\simeq$\,0.5}.

\begin{figure}[t]
\centering   
\includegraphics[scale=0.48, angle=0]{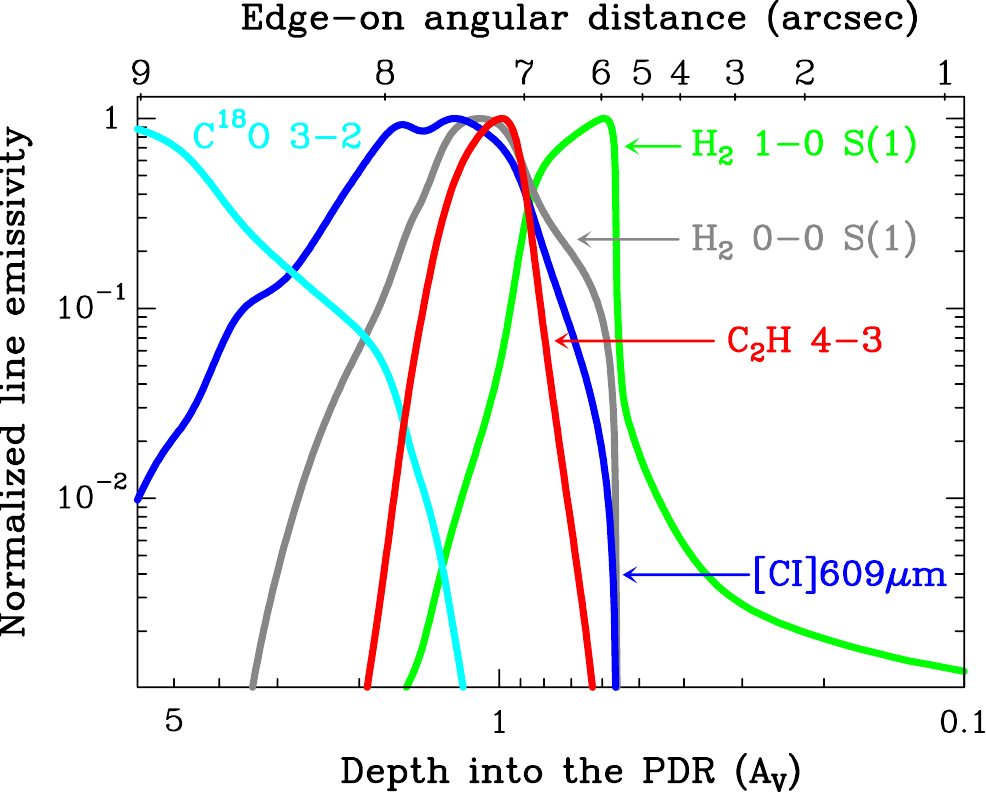}
\caption{Local line emissivities predicted by a PDR model with 
\mbox{$P_{\rm th}$/$k_{\rm B}$\,$=$\,10$^8$\,K\,cm$^{-3}$}. The upper horizontal axis
shows the equivalent angular scale for a perfectly edge-on PDR.
These distances will decrease as the inclination of the PDR with respect to a edge-on
PDR increases.}
\label{fig:PDR_line_emissivity}
\end{figure}

The model predicts steep temperature and density gradients
at small spatial scales, ranging from \mbox{$T_{\rm k}$\,$\simeq$\,700-600\,K} and 
$n_{\rm H}$ of a few 10$^5$\,cm$^{-3}$ at the hydrocarbon radical peak, to 
\mbox{$T_{\rm k}$\,$\simeq$\,150\,K} and \mbox{$n_{\rm H}\gtrsim$10$^6$\,cm$^{-3}$}
at the CO-rich  zone. \mbox{Figure~\ref{fig:PDR_line_emissivity}} shows that the predicted
 angular separation  between the local emissivities of 
\mbox{H$_2$ 0--0 $S$(1)}, C$_2$H $N$\,=\,4--3, and [\CI]\,609\,$\upmu$m  lines 
is very small, less than 1$''$ (for an edge-on PDR). This model  predicts that the
CO/C and H$_2$/H transition zones are separated by 0.0015\,pc, which
agrees with the observed emission stratification in DF3 \mbox{(Fig.~\ref{fig:crosscuts})}.
If \mbox{$P_{\rm th}$/$k_{\rm B}$} increases (decreases) by a factor of two, the predicted separation decreases (increases) by a factor of three.

\section{Discussion}
\label{sec:discussion}

\subsection{Far-UV-driven gas-phase hydrocarbon chemistry}
\label{subsec:gas_chemistry}

\begin{figure}[t]
\centering   
\includegraphics[scale=0.47, angle=0]{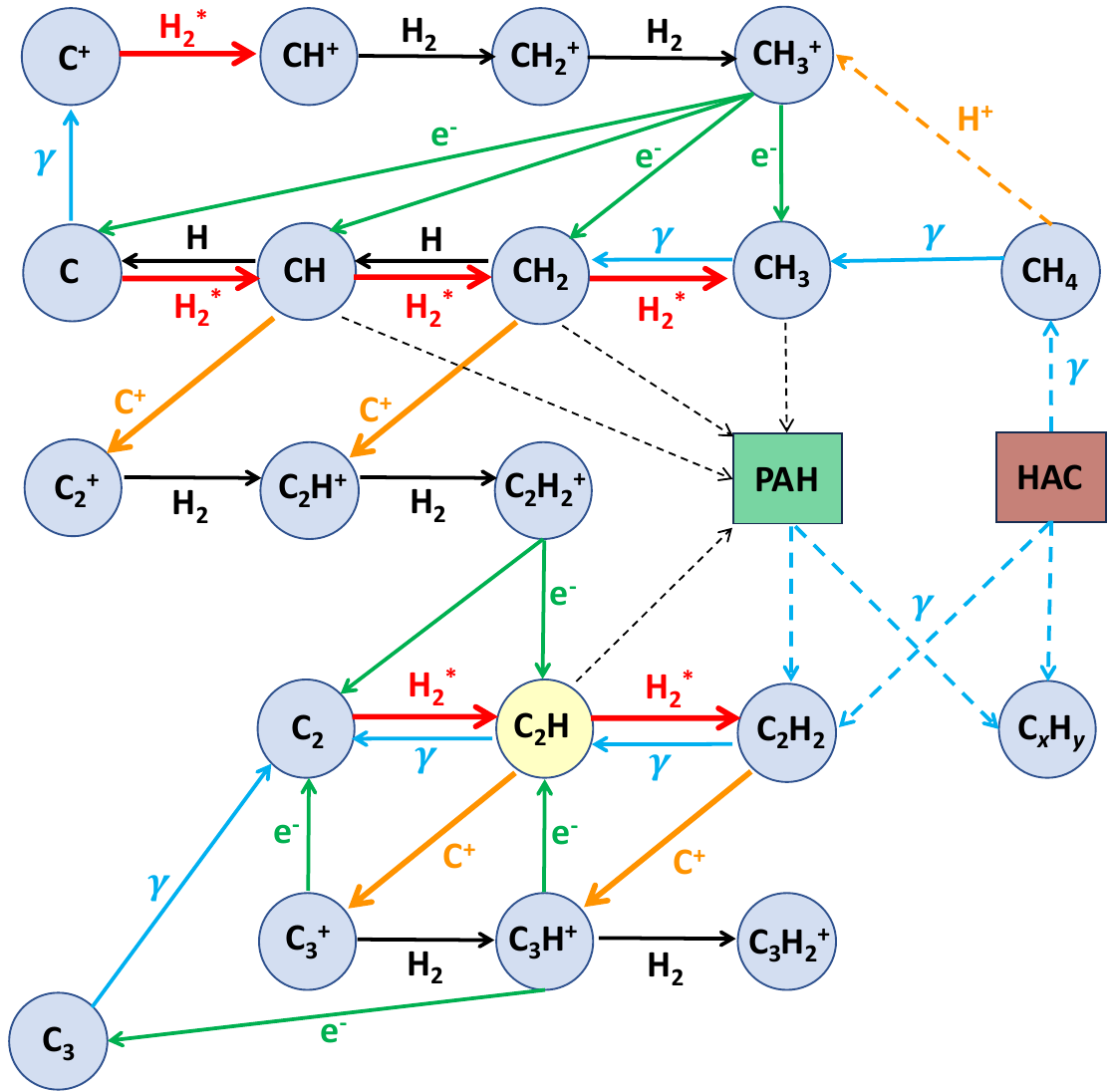}
\caption{Gas-phase formation and destruction pathways
  at the hydrocarbon radical peak (model in \mbox{Fig.~\ref{fig:PDR_model}}). 
Red arrows indicate endoergic reactions, which proceed rapidly at high $T$ or in regions with significant \mbox{FUV-pumped} H$_2^*$.
We also show possible reactions (dashed) involving  PAHs and HAC grains, which may be relevant in certain conditions.}
\label{fig:chemical_network}
\end{figure}

Figure~\ref{fig:chemical_network} summarizes the dominant gas-phase  reactions  of  hydrocarbons   in our reference model. 
The starting reaction is
 \begin{equation}
{\rm C^+}\,+\,{\rm H_2}(v,J)\,\rightarrow\,{\rm CH^+}\,+\,{\rm H},
\label{reaction_chp}
\end{equation}
 which has an endoergicity\footnote{Reaction~\ref{reaction_chp} becomes exoergic for H$_{2}^{*}$
 at
 \mbox{$v$\,$=$\,0, $J$\,$\geq$\,11}, and \mbox{$v$\,$=$\,1, $J$\,$\geq$\,7.} }
of $\Delta E$/$k$\,=\,4,300\,K  when H$_2$ is in the ground state $v$\,=\,0 \citep{Hierl97,Zanchet13b}. Thus,
this reaction is exceedingly slow and inefficient in cold  clouds
shielded from FUV radiation.
The warm temperatures and enhanced abundances of FUV-pumped  \mbox{H$_{2}^{*}$\,($v$$\geq$1)} in dense PDRs \citep[detected up to $v$\,=\,12 in the  Bar,][]{Kaplan21} overcomes the reaction endoergicity. This triggers the formation of 
abundant CH$^+$ \citep[e.g.,][]{Agundez10}, which peaks slightly ahead of the DF,
where \mbox{H$_{2}^{*}$\,($v$$\geq$1)} reaches its highest abundance. 
 CH$^+$ ro-vibrational line emission is readily detected along the  Bar 
\citep[][]{Naylor10,Nagy13,Parikka_2017,Zannese25}. Subsequent (fast) exothermic hydrogenation reactions
lead to the formation of CH$_{3}^{+}$ (the methyl cation), first detected in space by JWST in the irradiated disk 
\mbox{d203-506} \citep{Berne23} and also present in the Bar 
\citep{Zannese25}.  CH$_{3}^{+}$  reacts extremely slowly with H$_2$ 
due its very high endothermicity \citep[and no CH$_{4}^{+}$  products are observed in experiments;][]{Smith82,Asvany04} and is predicted to be
the most abundant hydrocarbon in the DF (\mbox{Fig.~\ref{fig:PDR_model}}).  CH$_{3}^{+}$
destruction is dominated by dissociative recombination, 
leading   to the formation of abundant radicals  CH$_2$ (methylene) and CH (methylidyne).
The slower CH$_{3}^{+}$ radiative association leads to somewhat lower levels of CH$_3$ (methyl).

Hydrogen abstraction reactions play a key role in the gas-phase growth of simple hydrocarbons. 
In \mbox{Fig.~\ref{fig:chemical_network}}, however, the reactions indicated
by  red arrows are endoergic.  Thus, they are very slow in cold  gas but, similarly to \mbox{reaction~(\ref{reaction_chp})}, become fast in FUV-irradiated gas due to the high temperatures and presence of \mbox{FUV-pumped} H$_{2}^{*}$. These reactions  boost the formation of CH, CH$_2$, and CH$_3$. These  radicals further react with C$^+$, promoting the formation
of hydrocarbons with two carbon atoms. CH is abundant in the  Bar \citep[and
correlates with the C$_2$H emission,][]{Nagy17} and across large scales 
in \mbox{OMC-1} \citep{Goico19}, following the    
spatial distribution of CH$^+$ and [\CII]\,158\,$\upmu$m  \citep{Goico19}.
This  suggests that reactions of CH and C$^+$ drive the formation of
 C$_{2}^{+}$, which starts the formation of hydrocarbons with two C atoms.

The enhanced abundances of CH$_{n}^{+}$ cations  and CH$_{n}$ radicals are a key feature of the carbon
chemistry in FUV-irradiated gas. Their abundances  
peak  ahead of the CO/C transition, implying that C$^+$  and H$_{2}^*$  are abundant, 
and increase with rising $P_{\rm th}$ due to corresponding increase in gas temperature, density, and 
\mbox{H$_{2}^*$ ($v \geq 1$)} abundances.
The C$_2$H abundance reaches a maximum at the DF (slightly behind the CH$^+$ peak).
At this maximum, C$_2$H formation is dominated by
the following reaction: 
\begin{equation}
{\rm C_2}\,+\,{\rm H_2}(v,J)\,\rightarrow\,{\rm C_2H}\,+\,{\rm H},
\label{reaction_C2_H2}
\end{equation}
which has an energy barrier\footnote{Since state-dependent rate constants,
$k_{v,J}$($T$), do not exist for \mbox{reactions 
\mbox{H$_2$($v$,$J$)\,+\,CH$_n$\,$\rightarrow$\,CH$_{n+1}$\,+\,H}
and for reaction~\ref{reaction_C2_H2}},
we modeled them by adopting state-dependent rate constants where the energy $E_{v,J}$ of each H$_{2}^{*}$ ro-vibrational state is
subtracted from the reaction endoergicity $\Delta E$ (when \mbox{$\Delta E > E_{v,J}$}). That is,  
\mbox{$k_{v,J}(T)$\,$\propto$\,exp\,$(-[\Delta E - E_{v,J}]/k_{\rm B}T)$}.} 
of $\Delta E$/$k$\,=\,1500\,K \citep{Pitts_1982}.
 On the other hand, destruction of C$_2$H is dominated
by photodissociation and by reactions with H$_2$ and with C$^+$. 
The reference model with \mbox{$P_{\rm th}$/$k_{\rm B}$$=$\,10$^8$\,K\,cm$^{-3}$} predicts a peak [C$_2$H] abundance of \,$\simeq$\,2.5$\times$10$^{-7}$, which 
agrees with the abundance derived from observations of DF3 and DF2
(\mbox{Sect.~\ref{sec:mtc_mods}}). The predicted\footnote{The (face-on PDR) column densities predicted by this model
are \mbox{$N$(C)\,=\,2$\times$10$^{17}$\,cm$^{-2}$} and \mbox{$N$(C$_2$H)\,=\,2$\times$10$^{14}$\,cm$^{-2}$}, a factor of $\sim$10 lower
than the observed values  
(Sects.~\ref{sec:mtc_mods} and ~\ref{sec:mtc_CI_mods}). This implies a geometrical intensity enhancement 
(sin\,$\alpha$)$^{-1}$\,$\simeq$\,10, corresponding  to a tilt angle of \mbox{$\alpha$\,$\approx$\,5$^o$},
as commonly found for the Bar \citep[e.g.,][]{Peeters24}.} \mbox{$N$(C)/$N$(C$_2$H)} column 
density ratio across the PDR, $\sim$10$^3$,  also matches the 
observed value \mbox{(Sect.~\ref{sec:mtc_CI_mods})}.
Lower \mbox{$P_{\rm th}$} values underestimates the C$_2$H abundances and line intensities \mbox{(Appendix~\ref{subsec:pdr2mtc})}.

Our observations and models demonstrate how FUV radiation triggers a specific
gas-phase hydrocarbon chemistry in dense PDR gas. As $G_0$ increases, so does the gas temperature and the  column density of C$^+$ and H$_{2}^{*}$. 
\mbox{Reaction~\ref{reaction_chp}} initiates this chemistry and 
represents the most relevant
destruction mechanism for C$^+$ at the DFs. This leads to the formation
of CO and HCO$^+$ close to the DFs 
\citep[as observed by ALMA;][]{Goico16}.
Furthermore, the observed extended spatial distribution of CH and \mbox{CH$^+$ $J$\,=\,1--0}  rotational emission in \mbox{OMC-1} \citep{Goico19}  probes  the widespread occurrence of hydrocarbon radical peaks across the   surfaces of \mbox{OMC-1},
including the Bar.

\begin{figure}[t] 
\centering   
\includegraphics[scale=0.42, angle=0]{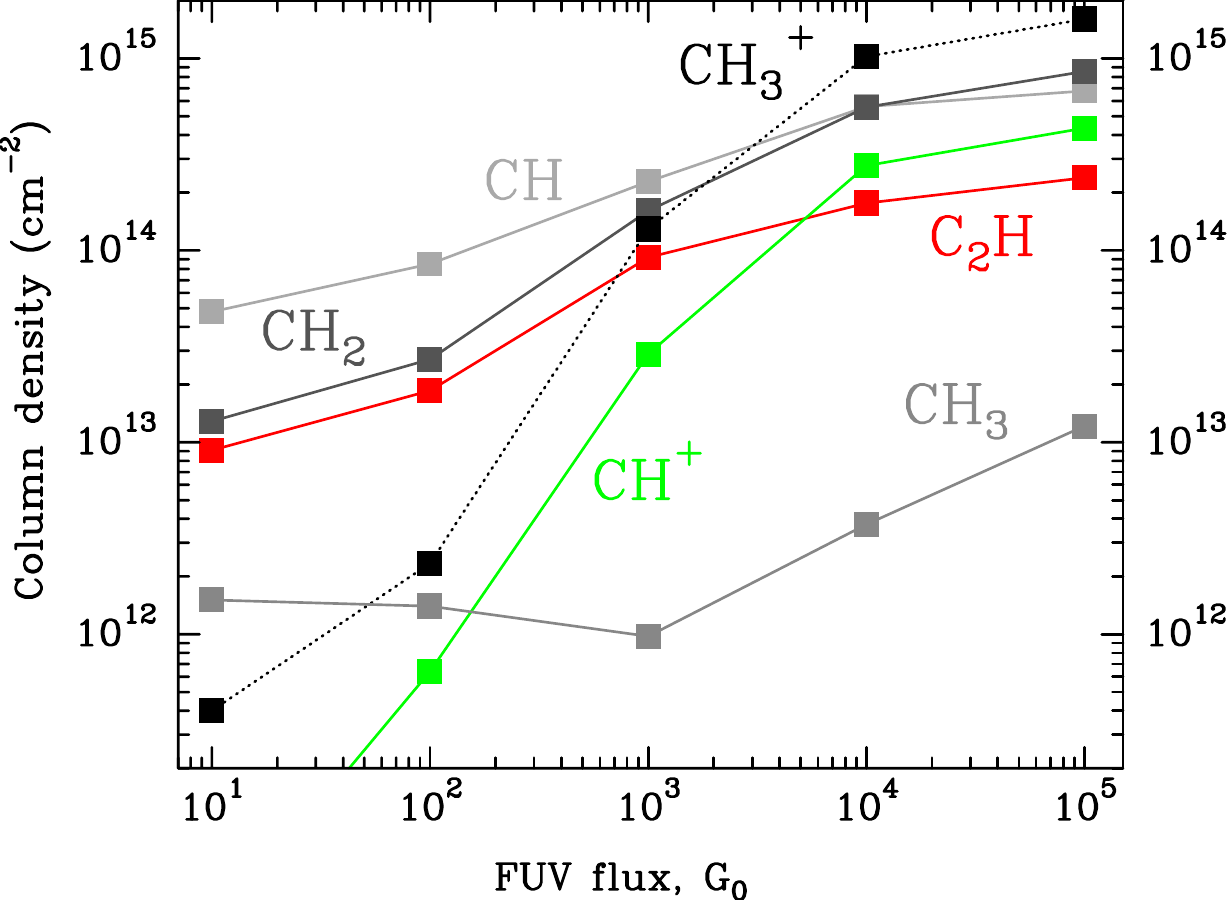}
\caption{Column densities of simple hydrocarbon radicals, CH$^+$,
and CH$_{3}^{+}$
as a function of increasing $G_0$. These models refer to
a  face-on PDR with $A_{V,\rm{tot}}$\,=\,10\,mag
and a constant density of $n_{\rm H}$\,=\,10$^5$\,cm$^{-3}$.}
\label{fig:columns}
\end{figure}

To isolate  the role of $G_0$ on this chemistry, \mbox{Fig.~\ref{fig:columns}} shows the predicted column densities of simple hydrocarbons as a function of  
$G_0$ for a PDR of constant density, $n_{\rm H}$\,=\,10$^{5}$\,cm$^{-3}$, representative
of $n_{\rm H}$ in the illuminated surfaces of \mbox{OMC-1} \citep[e.g.,][]{Pabst24}.
 \mbox{Figure~\ref{fig:columns}} shows that $N$(CH$^+$) increases by more than 3 orders of magnitude  from \mbox{$G_0$\,=\,10} to \mbox{10$^5$} \citep[see also,][]{Agundez10}. This enhancement triggers the formation of
related hydrocarbons when \mbox{$G_0$\,$>$\,10$^2$}. In these models, CH, CH$_2$, CH$_3$, and C$_2$H  increase
their column densities by factors of $\sim$\,15, $\sim$\,65, $\sim$10, and $\sim$\,25, respectively, compared to low FUV conditions.

Interestingly, the very high C$_2$H abundances inferred toward the molecular edge of the Bar are similar to those derived in galaxies undergoing vigorous star formation, where the origin of C$_2$H has been also attributed to gas with very low visual extinction \mbox{($A_V$\,$<$\,2\,mag)} in the form of thin, irradiated cloud interfaces \citep[][]{Burillo17}. In this context, the enhanced abundances of C$_2$H serve as a powerful tracer of radiative feedback.

\subsection{PAH and very small grain photodestruction as  sources of hydrocarbons in the atomic zone}

\label{sub:chem_ahead_DFs}

Our observations  show faint  C$_2$H emission\footnote{The brightest C$_2$H features in the atomic PDR zone (and with a line centroid  at $v_{\rm LSR}$\,$\simeq$\,10--11\,km\,s$^{-1}$) show \mbox{$I$(C$_2$H 4--3)\,$\simeq$\,5\,K\,km\,s$^{-1}$}, which translates into 
\mbox{[C$_2$H]\,$\lesssim$\,10$^{-8}$} (after non-LTE modeling and
adopting \mbox{$n_{\rm H}$\,$\simeq$\,(5--10)$\times$10$^{4}$\,cm$^{-3}$} and
\mbox{$T_{\rm k}$\,$\simeq$\,600--1000\,K}).
These values would represent the maximum C$_2$H 
abundance produced by \mbox{top-down} processes.} toward the atomic PDR zone (\mbox{Sect.~\ref{sec:channel maps}}). However, the gas-phase chemistry described in the previous section results in negligible levels of hydrocarbon radicals within these layers, which are characterized by low $A_V$ and \mbox{[H$_2$] $\ll$ [H]} (see \mbox{Fig.~\ref{fig:PDR_model}}).
Therefore, if this line emission originates from the Bar (as suggested
by its velocity centroid), not from background \mbox{OMC-1} layers, \mbox{top-down} formation processes must be considered; for example photo-processing of PAH, of small carbonaceous grains, or both.

 Only the  most resistant and stable population of PAHs with \mbox{$\gtrsim$\,50 C} atoms \citep[][]{Bakes01,Allain96,Andrews15,Montillaud13} are expected to survive under strong FUV irradiation conditions in the atomic PDR. \mbox{Indeed}, the observed evolution of the AIB emission and profiles across the Bar implies that  small PAHs,  aliphatic side groups, and very small carbonaceous grains undergo photodestruction in the atomic PDR \citep[][Schefter et al., in prep., Khan et al. in prep.]{Peeters24,Schroetter24,Chown24,Pasquini24}.
 
Laboratory experiments show that the photodestruction of PAHs
 produces  C$_2$H$_2$ \mbox{(acetylene)} as the main carbon-bearing fragment \citep[][]{Jochims94,Ekern98,Zhen15}.
Given that the photodissociation of C$_2$H$_2$ produces C$_2$H \citep[see  \mbox{Fig.~\ref{fig:chemical_network}};][]{Cheng11,Heays17}, this process may 
explain the presence of C$_2$H in the atomic PDR.
In addition,  experiments show that photodissociation of aliphatic PAHs  produce
hydrocarbons such as CH$_x$, C$_2$H$_x$, and C$_3$H$_x$ \citep[][]{Marciniak21}.
However, the aliphatic content of the AIB carriers in the atomic PDR is small, with the ratio of carbon atoms in aliphatic units to those in aromatic rings being approximately $<$\,1\%~(see the next section). 
 Finally, experiments also show that the photolysis of  hydrogenated amorphous carbon (HAC) grains leads to the production of a large variety of hydrocarbon molecules,  in particular \mbox{methane} (CH$_4$), and, to a lesser extent, C$_2$H$_x$ and  C$_3$H$_x$  
\mbox{\citep[e.g.,][]{Alata14,Duley15}}. 

Overall, photolysis of aliphatic PAHs, small PAHs, and HAC grains may inject hydrocarbons and temporarily increase their abundance in a PDR \citep[e.g.,][]{Awad22}. However, in a strongly irradiated PDR, the timescale of this photoprocessing is short ($\sim$10$^3$\,yr; \citealt{Jones14}), and the subsequent  photodissociation of any daughter molecular fragments much faster.
These time-scales contrast with the crossing time of
\mbox{$t_{\rm c}$\,$\simeq$\,3\,$\times$\,10$^4$}\,yr  for  material advected from the molecular PDR  to the atomic PDR,  at  
\mbox{$\sim$1\,km\,s$^{-1}$} in the Bar  \citep[][]{Pabst19}. 

To be more quantitative, 
\cite{Murga20} developed a  model of the Bar simulating the photoprocessing of
  PAHs. In Murga's model (their \mbox{Fig.~A.2}),  enhanced C$_2$H$_2$ abundances due to the photodestruction of PAHs occur in the atomic PDR  (\mbox{$A_V$\,$<$\,1 mag}), but over short timescales \mbox{($\sim$10$^4$\,yr\,$<$\,$t_{\rm c}$)}.
However, the resulting increase in C$_2$H abundance is small, likely  because  molecular fragments quickly photodissociate and most gas-phase carbon quickly converts into C and C$^+$.  
 \cite{Murga23} also developed  a time-dependent model of HAC photodestruction  (leading to CH$_4$ and other fragments) adapted to the strong FUV-illumination conditions of the Bar.
In their model, HAC photodestruction does not dominate either, and cannot explain the observed abundances of C$_2$H (this work) or of other small hydrocarbons detected by \cite{Cuadrado15}.
Only in the atomic PDR zone (at $A_V$\,$\simeq$\,0.1~mag) was this process  found to produce a  modest C$_2$H abundance \citep[$\lesssim$\,10$^{-10}$;][]{Murga23}. 

We conclude that photodestruction of PAHs or HAC grains may explain the transient presence of trace amounts of C$_2$H in the atomic PDR, assuming that PAHs and 
HAC grains are continually replenished   \citep[e.g., by evaporating very small grains,][]{Pilleri12} or quickly advected from the molecular cloud. These \mbox{top-down} processes 
may also contribute to the production of heavier hydrocarbons  (with more than two C atoms) elsewhere in the PDR \citep[e.g.,][]{Alata15}. 
However, 
our non-detection of C$_2$H$_2$ and CH$_4$ (the main hydrocarbon products of PAH and HAC grain photolysis) with JWST makes it difficult to predict their contribution.

\subsection{Links between hydrocarbon radicals and aliphatic PAHs.} 
\label{sub:link_with_aliphatic}

Previous observations revealed the decrease of the \mbox{3.4\,/\,3.3\,$\upmu$m} AIB intensity ratio with increasing FUV flux in PDRs \citep{Geballe89,Joblin96,Sloan97,Mori14,Pilleri15}.
This evolution is consistent with the photo-destruction of the more fragile bonds associated with the 3.4\,$\upmu$m band carriers \mbox{\citep[e.g.,][]{Marciniak21}}. 
With JWST, we spatially resolve the evolution of the \mbox{3.4\,/\,3.3\,$\upmu$m} AIB  ratio, which shows a particularly low value in the Orion Bar (a high $G_0$ PDR), ranging
from $\sim$\,0.1 in DF3 to $\sim$\,0.04 in the atomic PDR \citep{Peeters24,Chown24,Pasquini24}.
These values imply that the aliphatic component of the AIB carriers is small, comprising only $\sim$2\% (DF3) to $\sim$0.5\% (atomic PDR) of the carbon atoms in aliphatic groups compared to those in aromatic rings \citep[assuming plausible band strengths for the 3.3
and 3.4\,$\upmu$m bands; e.g.,][]{Yang13,Yang16}.

In the DFs, the spatial distribution of the C$_2$H  emission closely resembles the observed peaks of the 3.4/3.3~$\upmu$m AIB  ratio (\mbox{Figs.~\ref{fig:MIRI_images}} and \ref{fig:PDR_MIRI_crosscut}). 
This striking similarity may suggest a causal relationship between the
 abundance peaks of  hydrocarbon radicals and an increased proportion of PAHs with aliphatic side groups responsible for the $\sim$3.4~$\upmu$m AIB \citep[][]{Duley81,Muizon86,Jourdain90,Joblin96}.
Our PDR models show that the high abundance of C$_2$H 
(and that of  CH, CH$_2$, and CH$_3$ radicals) in the  DFs
can be attributed solely to gas-phase reactions initiated by C$^+$ and H$_{2}^{*}$.
In the DFs, the  abundance of these simple hydrocarbons  (all together)  is greater than 
the  abundance of typical PAHs containing $\sim$50~C atoms 
\citep[a few 10$^{-7}$ with respect to H nuclei; e.g.,][]{Tielens08}.
Therefore, it is conceivable that  highly reactive and abundant radicals react
in situ with PAHs, leading to PAHs with a small number of aliphatic side groups 
\citep[e.g., methylated PAHs;][]{Joblin96} and promoting the formation of additional aromatic rings \mbox{(e.g., reactions with C$_2$H)} which increases the size of the emitting PAHs. The recent detection of \mbox{CN-radical} derivatives of simple PAHs in \mbox{TMC-1}, such as naphthalene \citep[\mbox{C$_{10}$H$_8$};][]{McGuire21}, acenaphthylene \citep[\mbox{C$_{12}$H$_8$};][]{Cernicharo24}, and pyrene \citep[\mbox{C$_{16}$H$_{10}$};][]{Wenzel24a,Wenzel24b}, signals the importance of bottom-up gas-phase routes \citep[e.g.,][]{Kaiser15,Reizer22}, as CN radicals react readily with aromatic species \citep[e.g.,][]{Heitkmper22,Wenzel24b}.

Chemical and laboratory experiments  globally support this view
\mbox{\citep[e.g.,][]{Lemmens22}}  as indeed they demonstrate that PAHs  react with simple hydrocarbon radicals such as  
CH \citep[][]{Soorkia10,Goulay10,Reilly18,He20}, CH$_2$
 \citep[][]{Kraus93}, CH$_3$ \citep[][]{Shukla10,Zhao19,Levey22}, and 
 C$_2$H \citep[][]{Goulay06,Mebel08}.  
Some of these reactions may only be relevant
in high-temperature chemistry (e.g., the Bar), others 
can exhibit a positive temperature dependence but be efficient at low temperature  \mbox{\citep[e.g.,][]{Reizer22}}. 
We conclude that   reactions between simple but very abundant hydrocarbon radicals
and PAHs may locally contribute to increase the aliphatic content of the AIB carriers 
  (regardless of what the initial content was).
This bottom-up chemistry is then balanced, when PAHs are exposed to strong FUV fields, by photolysis of the aliphatic side groups, which reduces the  \mbox{3.4\,/\,3.3\,$\upmu$m} AIB ratio.

In dark clouds, most PAHs likely freeze out on grains due to their large binding energies and thus high condensation temperatures \citep[greater than that of water;][]{Piacentino2024}.
Experiments show that photolysis of PAHs in ices containing CH$_4$ result in methylation of these PAHs \mbox{\citep[][]{Bernstein2002}}.
In the Bar, the \mbox{3.4 / 3.3~$\mu$m} AIB ratio decreases again behind DF3 (\mbox{Fig.~\ref{fig:PDR_MIRI_crosscut}}), but much less steeply than the C$_2$H emission
 (which reflects the sharp reduction in gas-phase hydrocarbon  formation).
We hypothesize that the photodesorption of methylated PAHs from ices contributes to the increasing \mbox{3.4\,/\,3.3\,$\upmu$m} ratio in this deeper PDR zone. This may be a relevant mechanism in low-illumination PDRs, as they have colder grains covered by ices and a
less efficient gas-phase hydrocarbon production compared
to strongly irradiated PDRs \mbox{(Fig.~\ref{fig:columns}}).

 \section{Summary and conclusions}

We presented subarcsecond-resolution ALMA mosaics  of the Orion Bar PDR in  [\CI]\,609\,$\upmu$m ($^3$P$_1$--$^3$P$_0$), \mbox{C$_2$H ($N$\,=\,4--3)}, and \mbox{C$^{18}$O ($J$\,=\,3--2)} emission lines 
complemented by JWST \mbox{spectroscopic} images of H$_2$ and PAH emission.
We interpreted the data using up-to-date PDR and non-LTE radiative transfer models. 
We summarize our results as follows:
 
 -- The rim of the Bar shows a  corrugated, filamentary, and turbulent structure
  made of small-scale DFs that are bright and remarkably similar in both  IR H$_2$ and submillimeter  \mbox{C$_2$H 4--3}  emission. These fronts are engulfed in a PAH-emitting \mbox{halo} that separates the neutral, predominantly atomic edge of the PDR from the adjacent \HII~region.
The distribution of the \mbox{C$^{18}$O 3--2} emission is less filamentary but
clumpier, and it peaks deeper inside the  molecular cloud \mbox{(Sect.~\ref{sec:results})}.
The [\CI]\,609\,$\upmu$m emission peaks very close ($\lesssim$\,0.002\,pc) to the main DFs, suggesting  molecular gas structures with density gradients (\mbox{Sect.~\ref{sec:results}}).

--  The \mbox{C$_2$H $N$\,=\,4--3} emission traces hydrocarbon radical peaks in the DFs, 
slightly ahead of the CO/C transition zone. These peaks are characterized 
 by remarkably high C$_2$H abundances that reach up to several $\times$10$^{-7}$
relative to H. The C$_2$H emission profile more closely follows the IR H$_2$ emission than the gas temperature profile  \mbox{(Sect.~\ref{sec:analysis})}.

-- The high abundance of C$_2$H  (and related radicals CH$_3$, CH$_2$, and CH) at these peaks can be explained by simple gas-phase reactions driven by elevated temperatures, the presence of C$^+$  and C, and the  enhanced reactivity of FUV-pumped H$_{2}^{*}$
 \mbox{(Sect.~\ref{sec:pdr_mods})}.
This FUV-driven gas-phase carbon chemistry  is very efficient in dense PDR gas, with \mbox{$G_0$\,$>$\,10$^2$}
\mbox{(Sect.~\ref{subsec:gas_chemistry})}.

-- At low $A_V$, in the atomic PDR zone \mbox{(where [H]\,$\gg$\,[H$_2$])}, the AIB emission is the brightest, but 
aliphatic bonds and small PAHs are photo-destroyed \citep{Peeters24,Chown24,Schroetter24,Pasquini24}. Here, the production of  hydrocarbon radicals from gas-phase reactions is negligible. Thus, the detection of trace and transient amounts of C$_2$H may result from
top-down formation mechanisms, such as photoerosion of small PAHs and 
carbonaceous grains \mbox{(Sect.~\ref{sub:chem_ahead_DFs})}.

-- The C$_2$H emission  peaks coincide with the peaks of the 
\mbox{3.4/3.3\,$\upmu$m} AIB intensity ratio \mbox{(Sect.~\ref{sec:cut_C_H2})}, which is a proxy for the aliphatic-to-aromatic content of PAHs. 
This spatial coincidence implies that the conditions triggering the formation of simple hydrocarbons also favor the formation of PAHs with aliphatic side groups, 
potentially through \mbox{bottom-up} processes
 in which abundant CH$_n$ radicals react in situ with PAHs, locally enhancing their aliphatic content. Reactions of PAHs with heavier radicals, such as C$_2$H, may also promote the formation of additional rings, thereby increasing the size of the emitting PAHs \mbox{(Sect.~\ref{sub:link_with_aliphatic})}.

\vspace{0.1cm}

While this study highlights the role of gas-phase chemistry and suggests bottom-up processes in FUV-irradiated gas, similar observations of hydrocarbons containing three or more carbon atoms are needed to constrain the limits of this chemistry. 
In addition, a more precise assignment of the carriers of the 3.4\,$\upmu$m emission sub-bands across different PDR positions is needed to determine the relative contributions of methylated PAHs and \mbox{superhydrogenated} PAHs. 
The latter may be relevant in higher-density regions and could form within water ice mantles upon UV irradiation \citep{Bernstein99}.
 In general, combined observations of AIBs and C$_2$H (as a proxy for CH$_n$ radicals, whose rotational lines are found at less accessible wavelengths) in other PDRs 
 will be needed to draw more quantitative conclusions.

Interestingly, this carbon photochemistry is also relevant for 
planet-forming  disks affected by FUV \citep[e.g.,][]{Bosman21,Berne23,Goico24}, which  boosts the abundance of simple hydrocarbons. 
Our study calls for dynamical PDR models with chemical networks that gradually incorporate PAH formation, destruction, reactivity 
(e.g., with \mbox{radicals, atoms,
and  H$_{2}^{*}$}), and the photodesorption of frozen PAHs. Still, many reaction pathways  remain uncharacterized, making more theoretical and laboratory work necessary.

\begin{acknowledgements}  
We thank our referee for a concise but  constructive report. We made used of ADS/JAO.ALMA\#2021.1.01369.S data. ALMA is a partnership of ESO, NSF (USA) and NINS (Japan), together with NRC (Canada), NSTC and ASIAA (Taiwan), and KASI (Republic of Korea), in cooperation with the Republic of Chile. The Joint ALMA Observatory is operated by ESO, AUI/NRAO and NAOJ.
The JWST data were obtained from the Mikulski Archive for Space Telescopes at 
STSI, which is operated by the Association of Universities for Research in Astronomy, Inc., under NASA contract NAS 5-03127. Support for program \#1288 was provided by NASA through a grant from the Space Telescope Science Institute, which is operated by the Association of Universities for Research in Astronomy, Inc., under NASA contract NAS 5-03127.   JRG, SC, and MGSM thank the Spanish MCINN for funding support under grant
\mbox{PID2023-146667NB-I00}.
We thank the PCMI  of CNRS/INSU with INC/INP, co-funded by CEA and CNES.
EP acknowledges support from the University of Western Ontario, the Institute for Earth and Space Exploration, the Canadian Space Agency (CSA, 22JWGO1-16), and the Natural Sciences and Engineering Research Council of Canada. TO acknowledges the support by the Japan Society for the Promotion of Science KAKENHI Grant Number JP24K07087.
This project has received funding from the European Research Council (ERC) under the European Union’s Horizon Europe research and innovation programme ERC-AdG-2022 (GA No. 101096293).
MGSM acknowledges support from the NSF under grant CAREER 2142300.

\end{acknowledgements}

%
%

\bibliographystyle{aa}
\bibliography{references}

\onecolumn
\begin{appendix}\label{Sect:Appendix}

\section{Source coordinates}

Table~\ref{table:coordinates} shows the coordinates of the main DFs discussed in this work.

\begin{table}[h!]   
  \begin{center}
    \caption{Coordinates of main sources in the Orion Bar.}
   \begin{tabular}{l   c c c @{\vrule height 8pt depth 5pt width 0pt}}   
      \hline \hline
      Source     & $\alpha$(2000)     &   $\delta$(2000) &  Comment   \\ \hline   
    DF1          &      5:35:20.51      & $-$5:25:11.95  &  Dissociation front\\
   DF2          &       5:35:20.62      & $-$5:25:14.71  &  Dissociation front\\
    DF3          &      5:35:20.75      & $-$5:25:20.56  &  Dissociation front\\
    DF4          &      5:35:19.35      & $-$5:25:28.20  &  Dissociation front\\
    d203-506     &      5:35:20.32      & $-$5:25:05.55  &  Externally irradiated disk\\ 
\hline
\end{tabular}
\end{center}
\label{table:coordinates}
\end{table} 

\section{Beam dilution in single-dish observations: Beam coupling factors}
\label{App:beam_dilution}

\begin{table*}[h!]  
  \begin{center}
    \caption{Beam coupling corrections and C$_2$H line intensities toward the SDLS position (\mbox{Fig.~\ref{fig:RGB_Bar} left}).}
     \begin{tabular}{l   c c c c c c  c@{\vrule height 5pt depth 5pt width 0pt}}   
      \hline \hline
      Species     & Frequency  &    Telescope       &     HPBW       &                     &               &                & $I_{\rm corr}^{a}$  \\ 
                          & $[$GHz$]$  &   /\,Receiver      & $[$arcsec$]$   &  $f_b$(single-dish) &   $f_b$(ALMA) &  $f_b$(total)  & [K km\,s$^{-1}$]\\         
      \hline
C$_2$H~~$N=1-0$    &    87.3    & IRAM\,30m\,/\,E0  &    28.2        &           0.76               &   0.89            &   0.67         & 13.5 \\
C$_2$H~~$N=2-1$   &    174.7    & IRAM\,30m\,/\,E1  &    14.1        &           0.93               &   0.89                &   0.83         & 28.4\\
C$_2$H~~$N=3-2$   &    262.0    & IRAM\,30m\,/\,E2  &     9.4        &           0.99               &   0.89                &   0.88         & 38.7 \\
C$_2$H~~$N=4-3$   &    349.3    & IRAM\,30m\,/\,E3  &     8.0        &           1.00               &   0.89                &   0.89         & 37.0\\
C$_2$H~~$N=6-5$   &    524.0    & Herschel\,/\,HIFI &    40.5        &           0.66               &   0.89                &   0.59         & 13.7\\
C$_2$H~~$N=7-6$   &    611.3    & Herschel\,/\,HIFI &    34.7        &           0.70               &   0.89                &   0.62         & 8.7\\
C$_2$H~~$N=8-7$   &    698.5    & Herschel\,/\,HIFI &    30.4        &           0.74               &   0.89                &   0.66         & 4.7\\
C$_2$H~~$N=9-8$   &    785.8    & Herschel\,/\,HIFI &    27.0        &           0.77               &   0.89                &   0.68         & 3.6\\
C$_2$H~~$N=10-9$  &    873.1    & Herschel\,/\,HIFI &    24.3        &           0.80               &   0.89                &   0.71         & 3.6      \\
\hline
      \end{tabular}
  \tablefoot{In this work we correct the observed C$_2$H line intensities obtained
    with the IRAM\,30\,m and Herschel telescopes toward the SDLS position 
    as \mbox{$I_{\rm corr}$\,=\,$I_{\rm obs}$\,/\,$f_b$(total)}, with \mbox{$f_b$(total)\,=\,$f_b$(single-dish)$\cdot$}$f_b$(single-dish).
   $I_{\rm obs}$(IRAM) are taken from \cite{Cuadrado15}. 
   $I_{\rm obs}$(HIFI) are taken from Table A.1 of \cite{Nagy17}. $^a$Adding the intensities of 
   all individual fine-structure and hyperfine-structure C$_2$H lines.}
  \end{center}
  \label{table:f_b}
\end{table*} 

To approximately  correct for single-dish beam-size differences, we estimated a frequency-dependent \mbox{beam coupling factor} ($f_{\rm b}$)    using the spatial information provided by the high-angular resolution \mbox{C$_2$H $N$\,=\,4--3} map taken with ALMA.
 In doing this, we
assume that the emission from all C$_2$H rotational lines  have the same spatial distribution. 
We correct the observed  integrated line intensities, $I_{\rm obs}$, measured by the
IRAM\,30\,m and Herschel telescopes as: 
\mbox{$I_{\rm corr}$\,=\,$I_{\rm obs} / f_{\rm b}$}. 
We do this in a two step procedure.
We first spatially smoothed the large  \mbox{C$_2$H $N$\,=\,4--3}  map obtained with the IRAM\,30\,m telescope  to the different 
full width at half maximum (FWHM) beam  at the frequency of each C$_2$H  rotational line observed in the IRAM\,30\,m and Herschel/HIFI line surveys toward the SDLS position. We then compute
 \mbox{$f_{\rm b}$(single-dish)\,=\,$I_{\rm smooth}$(HPBW)/$I_{\rm obs}$(8$''$)}, where
 \mbox{$I_{\rm smooth}$(HPBW)} is the  
 intensity (in K\,km\,s$^{-1}$) of the \mbox{C$_2$H $N$\,=\,4--3} line extracted from the spatially smoothed maps toward the SDLS position. In a second step, we smoothed the high angular resolution ALMA C$_2$H $N$\,=\,4--3 map  to a 8$''$ angular resolution and compute
 \mbox{$f_{\rm b}$(ALMA)\,=\,$I_{\rm smooth}$(8$''$)/$I_{\rm obs}$(ALMA)}, where
 \mbox{$I_{\rm smooth}$(8$''$)} is the intensity
 of the  \mbox{C$_2$H $N$\,=\,4--3}  line 
  extracted from the smoothed ALMA maps toward the SDLS position.
  The final beam coupling correction factor is 
   \mbox{$f_{\rm b}$\,=\,$f_{\rm b}$(single-dish)\,$\cdot$$f_{\rm b}$(ALMA)}. The
 resulting 
  correction factors are listed in \mbox{Table~\ref{table:f_b}.1}.

\FloatBarrier 

\clearpage

\section{Complementary  observational figures}

\begin{figure}[htb]
\centering   
\includegraphics[scale=0.55, angle=0]{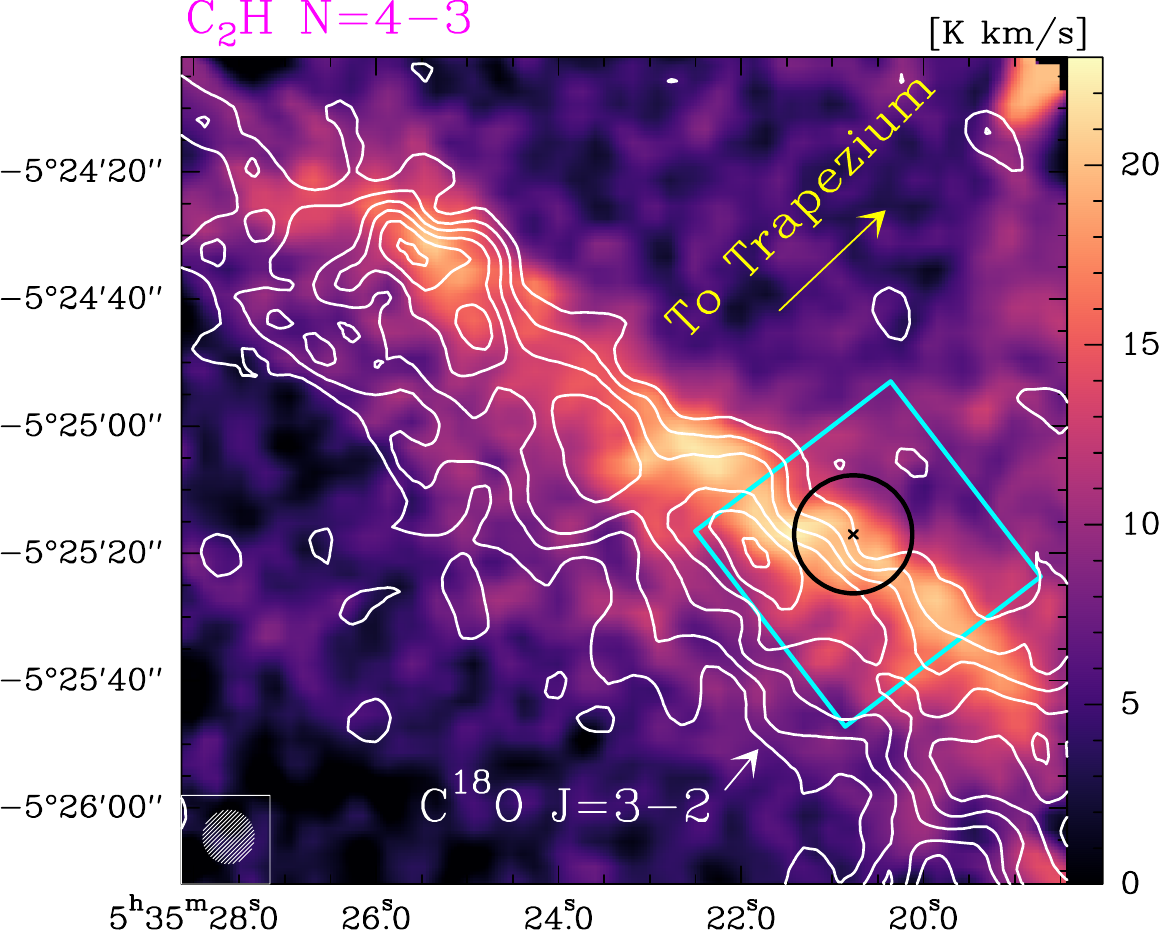}
\caption{IRAM\,30m map of the entire Orion Bar in the \mbox{C$_2$H 4--3}  integrated line emission at $\sim$8$''$ resolution (in R.A. and DEC. coordinates). White contours show the \mbox{C$^{18}$O 3--2}  emission
from 10 to 35~K\,km\,s$^{-1}$ in steps of 5~K\,km\,s$^{-1}$. 
The cyan square
shows the FoV observed with ALMA. The black circle shows the 
single-dish line survey position \citep[SDLS,][]{Cuadrado15} including the FoV
observed with JWST.}
\label{fig:map_e330}
\end{figure}

In this appendix, we present additional images that aim to more clearly demonstrate the relationships between the emission from different species across the  Bar.
\mbox{Figure~\ref{fig:map_e330}} shows a large view of the entire Bar as seen in
the \mbox{C$_2$H $N$\,=\,4--3} emission mapped with the IRAM\,30m telescope.
This emission traces an extended ridge of multiple DFs along the Bar.
White contours show the \mbox{C$^{18}$O 3--2}  emission tracing the more FUV-shielded
molecular  cloud and dense clumps.

\begin{figure}[htb]
\centering   
\includegraphics[scale=0.6, angle=0]{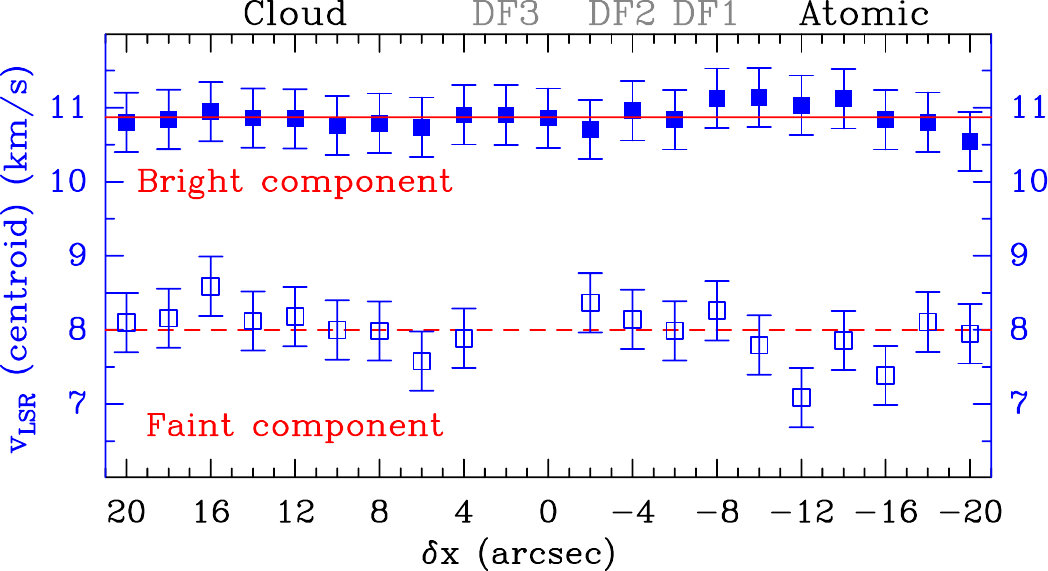}
\caption{Line emission velocity-centroids  of [\CI]\,609\,$\upmu$m obtained from
a two-Gaussian fit across cut~A. See the ALMA spectra in 
\mbox{Fig.~\ref{fig:spectra_cutA}}.}
\label{fig:cut_A_centroids}
\end{figure}

Figure~\ref{fig:cut_A_centroids}  [\CI]\,609\,$\upmu$m emission velocity-centroids obtained from a two-Gaussian fit across cut A (ALMA observations).

\begin{figure*}[htb]
\centering   
\includegraphics[scale=0.5, angle=0]{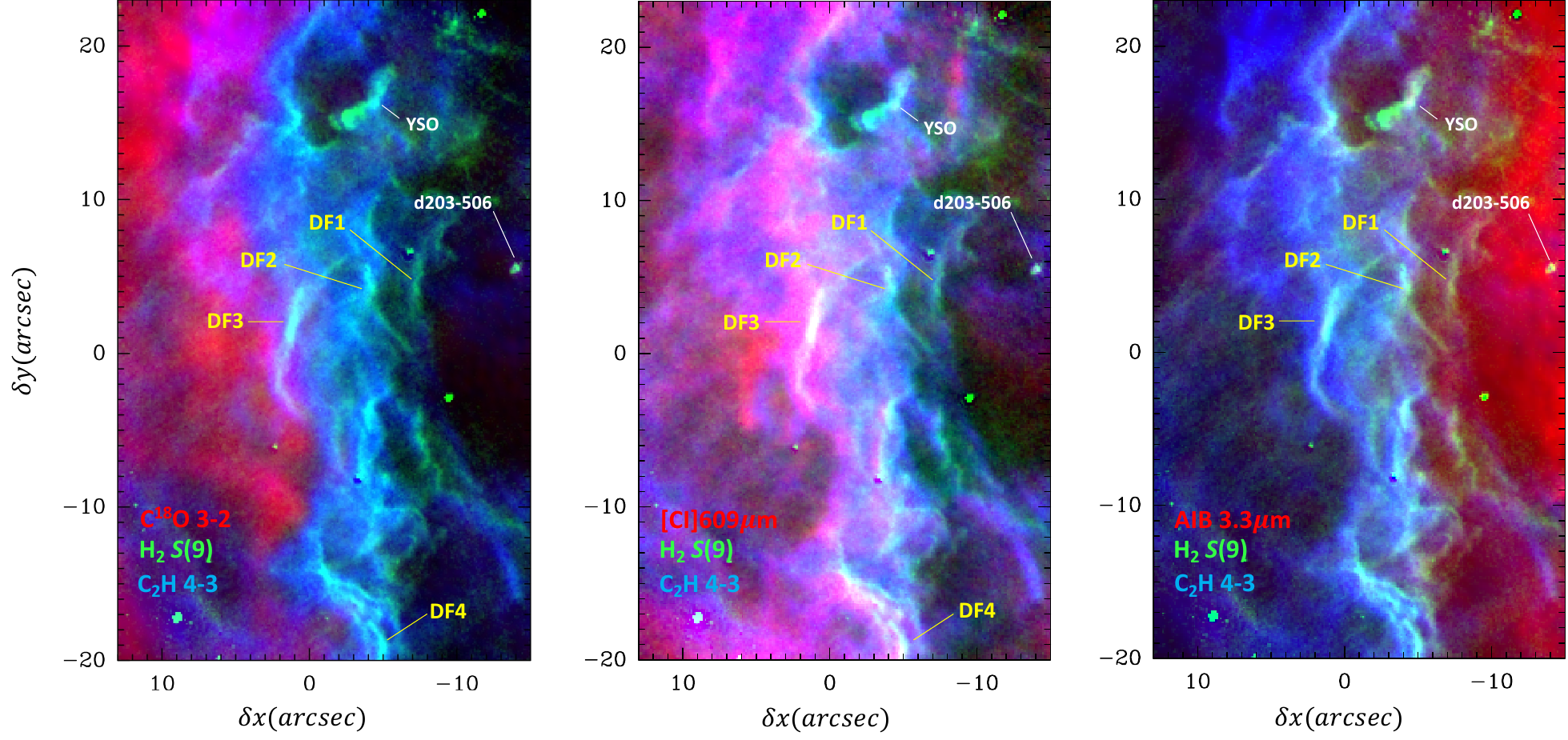}
\caption{Subarcsecond resolution RGB images of the Bar. 
Green represents the \mbox{NIRCam F470M$-$F480M} image 
(a proxy of the H$_2$ $v$\,=\,0--0 $S$(9) emission) and blue represents the
\mbox{C$_2$H $N$\,=\,4--3}, \mbox{$F$\,=\,5--4 and $F$\,=\,4--3} emission observed
with ALMA. From left to right, red represents \mbox{C$^{18}$O $J$\,=\,3--2} 
and [\CI]\,609\,$\upmu$m observed with ALMA, and JWST/NIRCam \mbox{F335M$-$F330M} image (PAH emission), respectively. We rotated the original images by 37.5$^o$ clockwise  to bring the FUV illumination from the Trapezium in the horizontal direction (from the right).}
\label{fig:contours}
\end{figure*}

Figure~\ref{fig:contours} shows   images of the edge of the Bar
combining ALMA and JWST observations. These images 
  reveal the small-scale structure and chemical stratification of the PDR.

\begin{figure*}[h]
\centering   
\includegraphics[scale=0.5, angle=0]{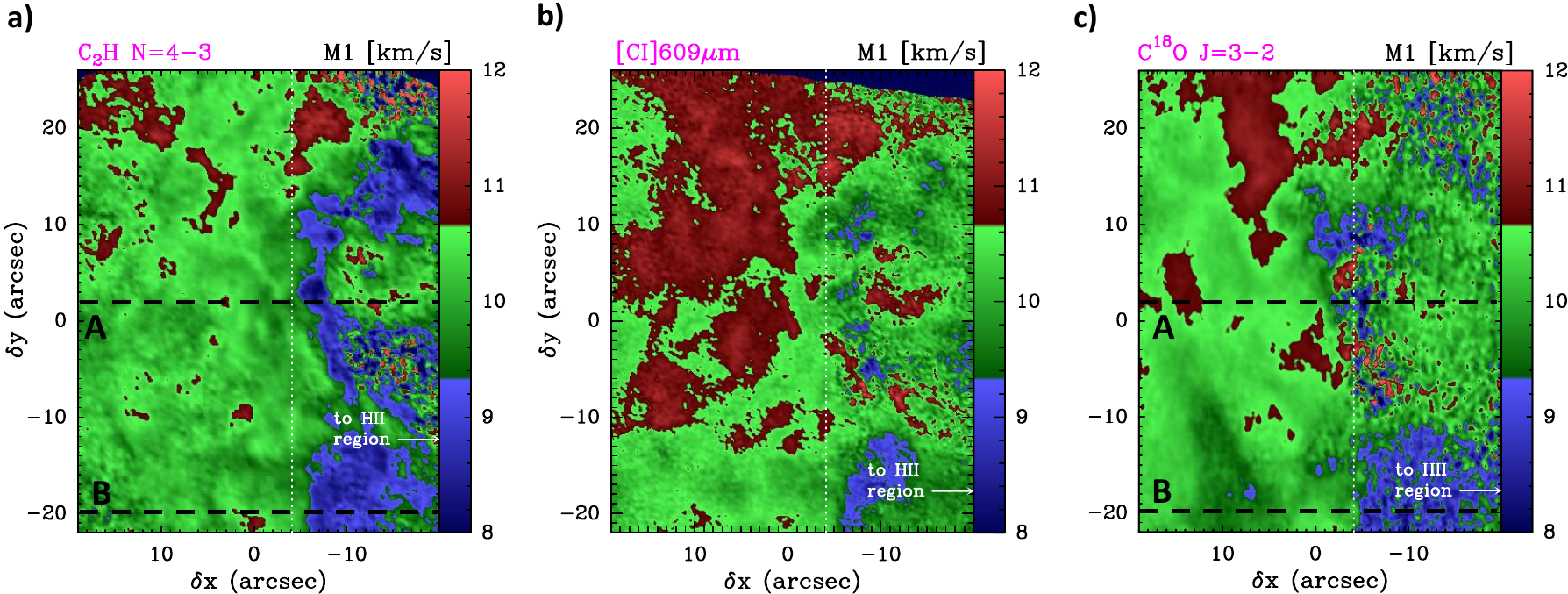}
\caption{Intensity-weighted mean LSR velocity maps (moment~1):
(a) C$_2$H $N$\,=\,4--3, (b) [\CI]\,609\,$\upmu$m, and (c) C$^{18}$O $J$\,=\,3--2.
The green shaded areas show LSR velocities consistent with emission from the Bar. The blueish points
(blue-shifted with respect to the Bar) show emission with more relevant contribution from  \mbox{OMC-1} in the background (e.g., DF1). The reddish points show red-shifted emission from the main velocities of the Bar PDR.
We rotated the original images by 37.5$^o$ clockwise  to bring the FUV illuminationin the horizontal direction (from the right).}
\label{fig:alma_moments}
\end{figure*}

\mbox{Figure~\ref{fig:alma_moments}} shows intensity-weighted mean LSR velocity
(moment~1) maps 
of the \mbox{C$_2$H $N$\,=\,4--3}, [\CI]\,609\,$\upmu$m, and \mbox{C$^{18}$O $J$\,=\,3--2}
lines observed with ALMA. The color code is such that the greenish areas are consistent with the main emission velocities arising from the Bar PDR, reddish areas reveal slightly redshifted emission (linked to the specific gas kinematics of the PDR), whereas
bluish regions mostly represent molecular gas in the background \mbox{OMC-1} cloud.
Toward the atomic PDR zone (to the right of the dashed vertical white line) the greenish
areas may also represent gas at deeper layers of \mbox{OMC-1} illuminated from a slanted angle.

\begin{figure*}[bh]
\centering   
\includegraphics[scale=0.50, angle=0]{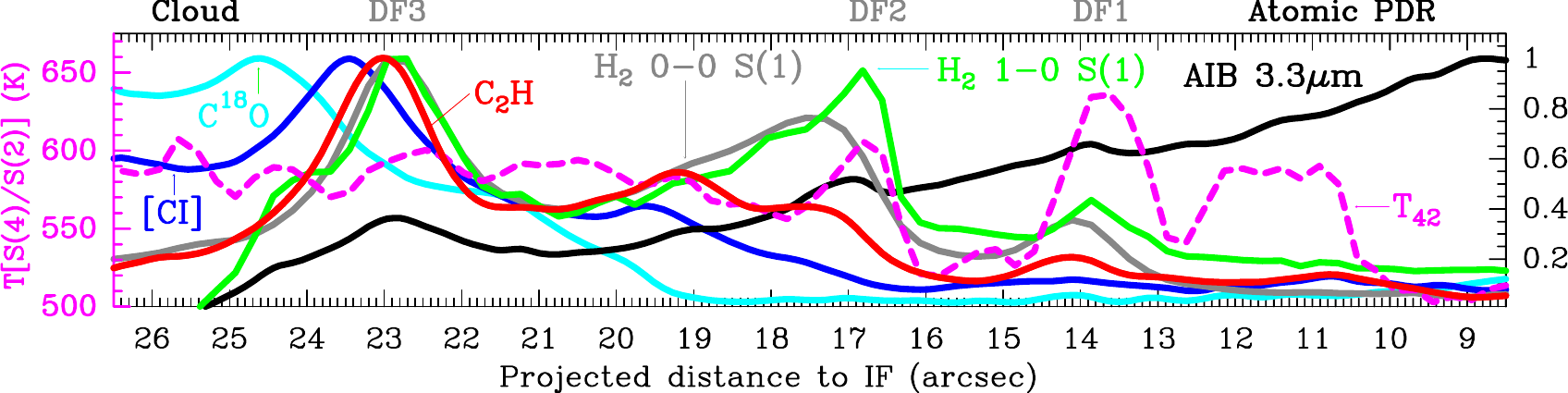}
\caption{Crosscut C: Vertically averaged  intensity ratio with \mbox{$\Delta(\delta y)$\,=\,2$''$}  crossing the green cross   in \mbox{Fig~\ref{fig:MIRI_images}}.
The  dashed line shows the rotational temperature $T_{\rm 64}$ (scale in the left y-axis)
derived from the $p$-H$_2$ $v$\,=\,0–0 $S$(4)/$S$(2) line intensity ratio observed with MIRI-MRS
(see text).}
\label{fig:cut_c_T42}
\end{figure*}

Figure~\ref{fig:cut_c_T42} shows cross cut C from observations with ALMA, MIRI-MRS, and NIRSpec spectrometers. This cut crosses DF1, DF2, and DF3 (similar to \mbox{Fig.~\ref{fig:PDR_MIRI_crosscut})}. The dashed magenta  curve shows 
the rotational temperature $T_{64}$ obtained from H$_2$ \mbox{$v$\,=\,0--0} $S$(4) and $S$(2) line intensities observed
with MIRI-MRS. Because of the relatively high gas densities at the DFs, $T_{64}$\,$\simeq$\,$T_{\rm k}$.

\newpage

\clearpage

\section{Line intensity correlations}\label{sec:correlations}

\begin{figure}[h]
\centering   
\includegraphics[scale=0.148, angle=0]{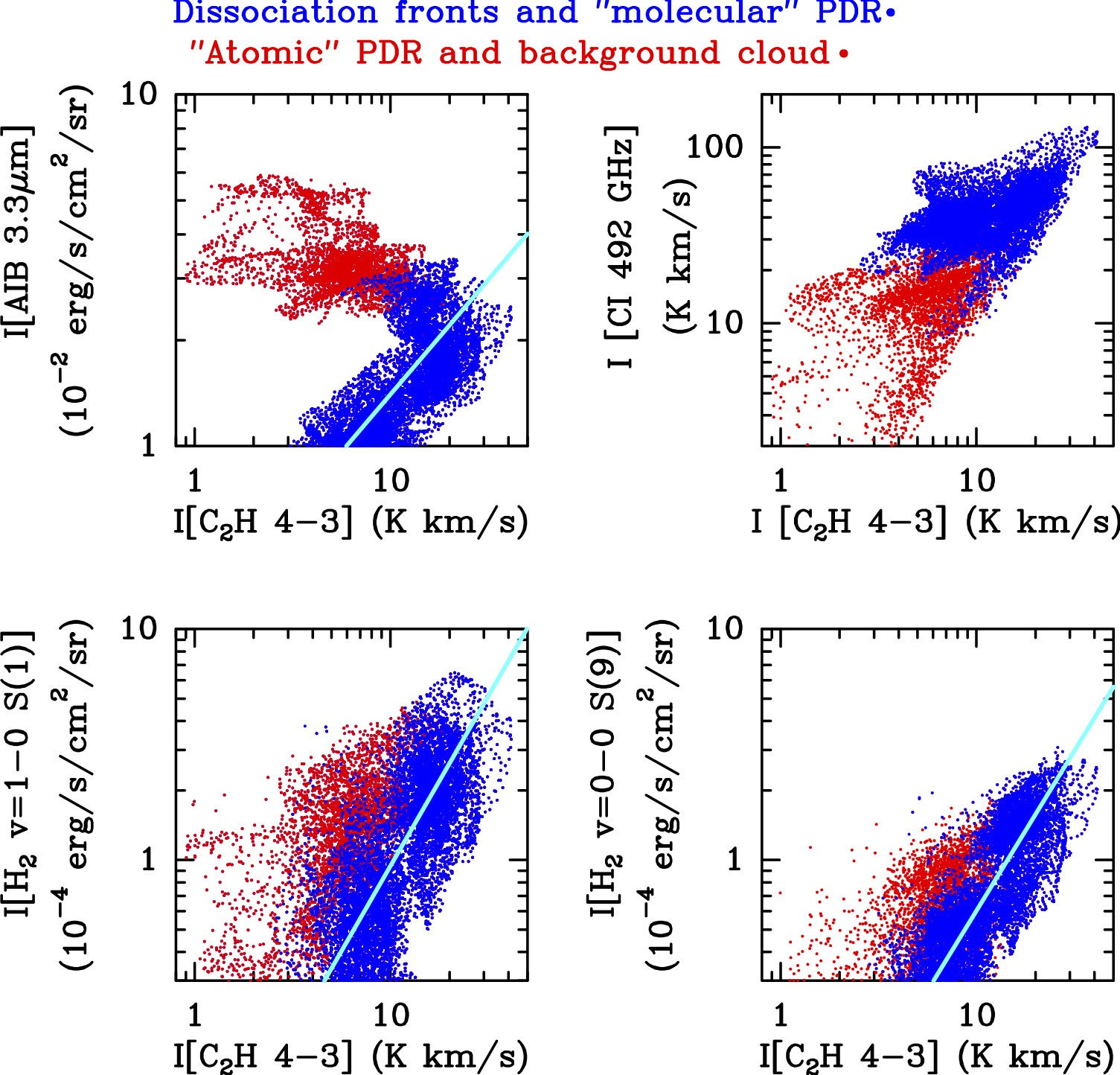}
\caption{Intensity correlation plots of C$_2$H $N$\,=\,4--3. Blueish pixels correspond 
to emission at $\delta$$x$\,$>$\,$-$5$''$ (main DFs and molecular PDR) whereas
reddish pixels correspond to emission at $\delta$$x$\,$<$\,$-$5$''$
(atomic PDR and background OMC-1 cloud). The straight cyan lines are
regression curves, with parameters of Table~\ref{table:correlations}, obtained by fitting the bluish areas only.} 
\label{fig:PDR_correlations}
\end{figure}

To support our discussion and PDR modeling results (namely, that C$_2$H formation in the DFs is dominated
by gas-phase chemistry initiated by reactions of C$^+$ with H$_{2}^{*}$), we constructed line-intensity correlation plots  from the larger 
NIRCam and Keck filter images shown in Fig.~\ref{fig:cch_c18o_jwst_integrated_intensity}
(using their common \mbox{30$''$\,$\times$\,30$''$ FoV}, \mbox{$\delta x$\,=\,[$+$15$''$,$-$15$''$]} and \mbox{$\delta y$\,=\,[$-$16$''$,$+$14$''$]}). 
We first spatially smoothed the images
to a common angular resolution of 0.8$''$.
\mbox{Figure~\ref{fig:PDR_correlations}} shows the resulting correlation 
plots\footnote{The IR images used in this correlation analysis  are continuum-subtracted NIRCam and Keck filter images \citep{Habart23a,Habart24}. They represent the intensity of the AIB 3.3\,$\upmu$m, \mbox{H$_2$ $v$\,=\,1-0 $S(1)$},
and \mbox{H$_2$ $v$\,=\,0-0 $S(9)$} emission with an accuracy of $\lesssim$\,20\,\%, depending on the position and environment \citep{Chown25}. This uncertainty contributes to the scatter seen  in the correlation plots \mbox{(Fig.~\ref{fig:PDR_correlations})}.}
after clipping intensities below a 3$\sigma$ detection threshold.
Blueish points show the measured line intensities in the molecular PDR, at  $\delta x$\,$>$\,$-$5$''$ (i.e., in the DFs and deeper into the molecular cloud). The reddish
points refer to  the atomic PDR zone (i.e., at $\delta x$\,$<$\,$-$5$''$).

Table~\ref{table:correlations} summarizes  power-law fits  for each correlation plot toward the molecular PDR (only taking the data in blue). To first order, the H$_2$, AIB\,3.3\,$\upmu$m, and [\CI]\,609\,$\upmu$m intensities monotonically  follow that
of   \mbox{C$_2$H $N$\,=\,4--3}  (as measured by the  \mbox{Spearman's} rank coefficient). 
The  \mbox{$I$(C$_2$H 4--3)--$I$(H$_{2}^{*}$)} intensities are the more linearly correlated ones (where the H$_2$ $v$=1--0 $S$(1) and $v$=0--0 $S$(9) excited lines mostly trace
FUV-pumped H$_{2}^{*}$).

The \mbox{$I$(C$_2$H\,4--3)--$I$(\CI\,609\,$\upmu$m)} plot shows that these emissions are also related, with a correlation trend that approximately extends into the atomic PDR
lines-of-sight (reddish areas).  That is, wherever these line intensities
come from (atomic PDR or background \mbox{OMC-1}), they are related. Finally, the  \mbox{$I$(C$_2$H\,4--3)--$I$(AIB 3.3\,$\upmu$m)}  plot shows a different behavior, with
two correlation trends: moderately correlated toward the molecular PDR (where the AIB emission
is fainter), and anti-correlated toward
the atomic PDR zone (where the AIB 3.3\,$\upmu$m band emission reaches  maximum intensity values).

\begin{table}[h]
\caption{Summary of the line intensity correlation plots  in Fig.~\ref{fig:PDR_correlations}.\label{table:correlations}} 
\centering
\begin{tabular}{llcccc@{\vrule height 8pt depth 5pt width 0pt}}
\hline\hline
$x$          &  $y$                                          &  $a$ &     $b$       & $\rho_P$  & $\rho_S$    \\ 
\hline
$I_{\rm C_2H\,\,4-3}$   & 10$^2$$\cdot$$I_{\rm AIB\,3.3\,\mu m}$    & 0.65 & --0.51  &  0.62      &  0.65     \\
$I_{\rm C_2H\,\,4-3}$   & $I_{\rm [CI]\,609\,\mu m}$                & 0.67 & 0.92    &  0.54      &  0.54     \\
$I_{\rm C_2H\,\,4-3}$   & 10$^4$$\cdot$$I_{\rm H_2\,0-0\,S(9)}$     & 1.39 & --1.61  &  0.76      &  0.79     \\
$I_{\rm C_2H\,\,4-3}$   & 10$^4$$\cdot$$I_{\rm H_2\,1-0\,S(1)}$     & 1.47 & --1.50  &  0.71      &  0.71     \\
\hline                                    
\end{tabular}
\tablefoot{Regressions fitted to the molecular PDR data (bluish areas in Fig.~\ref{fig:PDR_correlations}) in a weighted least-square fit,
\mbox{log$_{10}$\,$y$\,=\,$a$\,log$_{10}$\,$x$\,+\,$b$}. Coefficient $\rho_P$ is the Pearson coefficient
that measures the linearity of the log--log correlation. $\rho_S$ is the Spearman coefficient that
measures the monotonic relationship. The typical error in the fits for 
$a$ and $b$ is 0.01.}
\end{table}

\begin{figure}[h]
\centering   
\includegraphics[scale=0.55, angle=0]{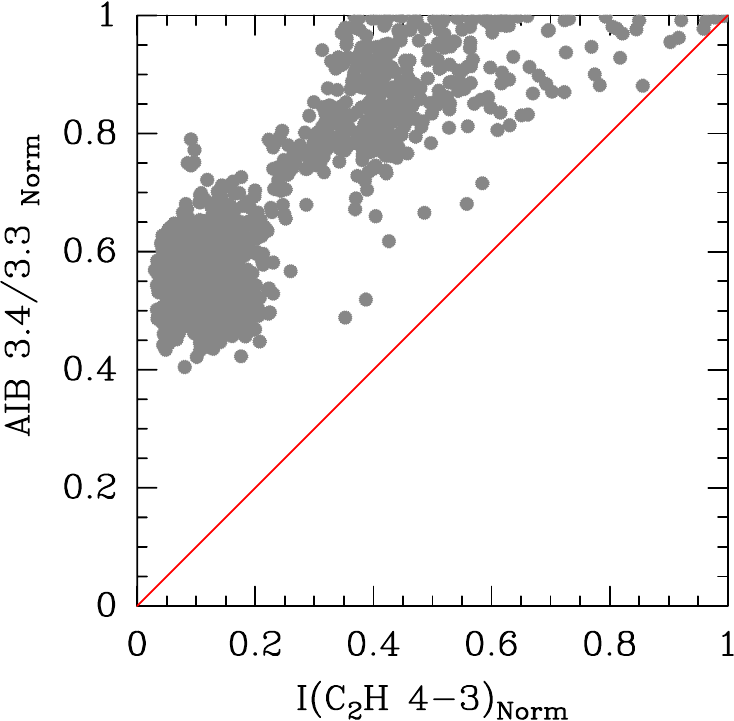}
\caption{Normalized 3.4/3.3\,$\upmu$m AIB ratio versus the normalized \mbox{C$_2$H 4--3} line intensity ratio
extracted from the small FoV in Fig.~\ref{fig:MIRI_images}.}
\label{fig:corr_34_33_C2H}
\end{figure}

\mbox{Figure~\ref{fig:corr_34_33_C2H}} shows 
the normalized 3.4/3.3\,$\upmu$m AIB ratio versus normalized \mbox{C$_2$H 4--3} line intensity ratio extracted from the small FoV shown in \mbox{Fig.~\ref{fig:MIRI_images}}
(the JWST spectroscopy FoV). This plot is a  proxy for the increasing aliphatic content of the AIB carriers relative to the increasing column density of simple hydrocarbon radicals. The red line shows a 1:1 linear correlation.

\section{Nonlocal and non-LTE collisional and radiative excitation of C$_2$H rotational lines}\label{App:C2H_collisonal}

In our nonlocal and non-LTE excitation and radiative transfer models \citep{Goico22}, we adopted the  fine-structure resolved   \mbox{C$_2$H--H$_2$} rate\footnote{These close-coupling quantum scattering calculations provide \mbox{C$_2$H--$o$/$p$-H$_2$} inelastic rate coefficients ($\gamma_{\rm H_2}$ in cm$^{3}$\,s$^{-1}$) up to
$T_{\rm k}$\,=\,500\,K. For higher gas temperatures, we  simply extrapolate these
rates as $\sqrt{T_{\rm k}}$. We assume that the H$_2$ OTPR is thermalized to $T_{\rm k}$, e.g., $\sim$1.6 at 100~K and $\sim$2.9 at 200~K.}
coefficients, $\gamma_{\rm H_2}$, 
recently computed by \mbox{\citet{Pirlot23}} up to C$_2$H rotational level \mbox{$N$\,=\,20}
($E_{\rm up}$/$k$\,=\,879\,K).
Hyperfine-resolved \mbox{C$_2$H--$e^-$} \citep{Nagy15} and \mbox{C$_2$H--H$_2$} collisional rate coefficients also exist \citep[but the latter only up to 100\,K;][]{Pirlot23}. We checked that for $T_{\rm k}$\,$\leq$\,100\,K, both datasets (fine-structure versus hyperfine-structure) produce similar results. In addition, since  \mbox{$\gamma_{\rm H_2}$\,$\cdot$\,$n$(H$_2$)\,$\gg$\,$\gamma_{\rm e}$\,$\cdot$\, $n_{e}$},
electron  excitation plays negligible role for C$_2$H  even for the highest possible
$e^-$ abundances near the DFs, $n_e$\,$\simeq$\,$n$(C$^+$)\,$\simeq$10$^{-4}$\,$n_{\rm H}$ \citep[][]{Cuadrado19,Pabst24}.

\begin{figure}[ht]
\centering   
\includegraphics[scale=0.42, angle=0]{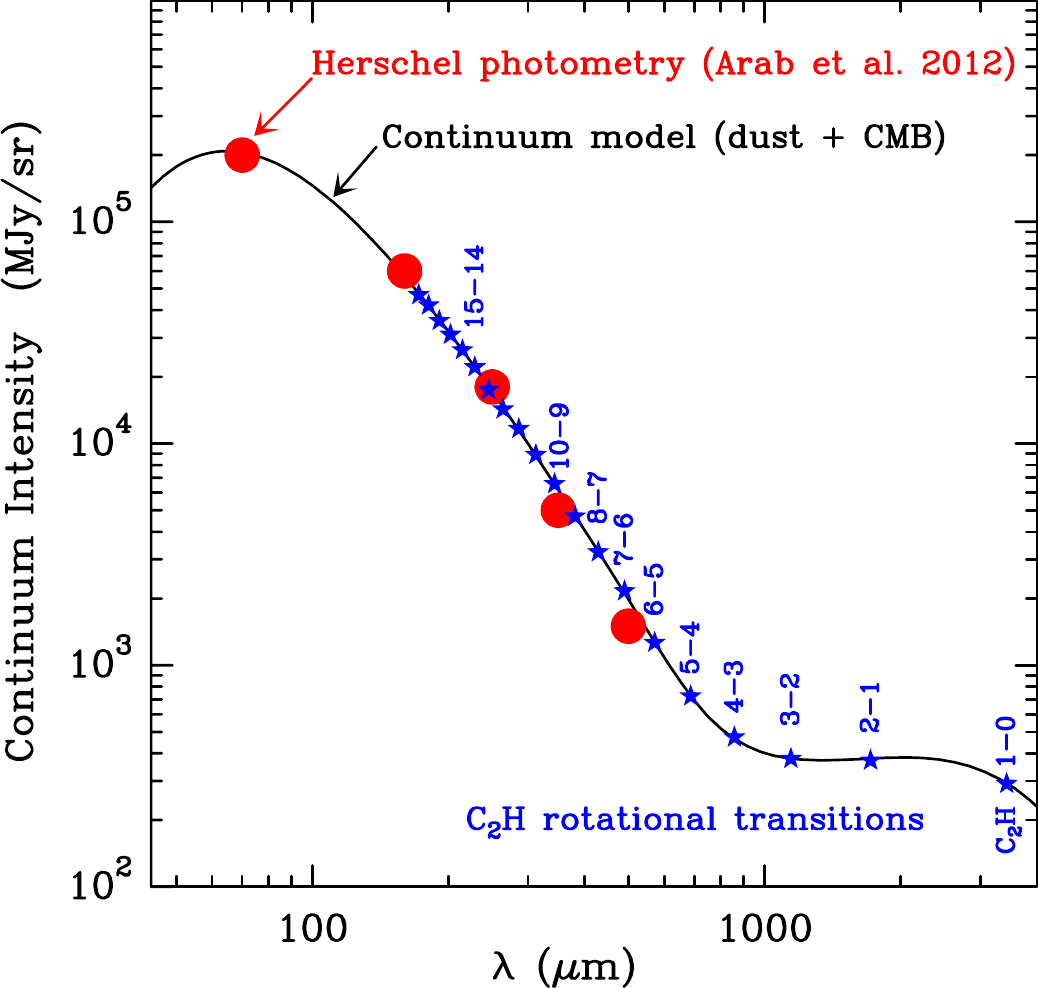}
\caption{Continuum emission model (a modified black body with $T_{\rm d}$\,=\,50\,K and the cosmic millimeter background) used in our C$_2$H excitation and radiative
transfer calculations. Red circles show Herschel’s photometric measurements in the Bar \citep{Arab12}. 
Blue stars show the wavelength position of the C$_2$H rotational
transitions.}
\label{fig:SED}
\end{figure}

In our excitation models, we include radiative excitation by absorption of dust continuum photons. We approximate
this far-IR and submillimeter continuum with a modified blackbody, with a color temperature
of 50~K and a wavelength-dependent (in $\upmu$m) dust continuum opacity that varies as 0.03(160/$\lambda$)$^{1.6}$.
\mbox{Figure~\ref{fig:SED}} shows the synthetic  continuum emission (dust and cosmic microwave background) used in our radiative transfer model. This is the continuum (external radiation and internally generated field) felt by the molecular gas in the Bar. The red dots show
Herschel's continuum measurements \citep{Arab12}. The blue stars show the wavelength
position of C$_2$H rotational lines.

  Our best models of the C$_2$H emission in the DFs, predict  a rotational excitation temperature of \mbox{$T_{\rm rot}$(C$_2$H 4--3)\,=\,20--24 K} (subthermal excitation), in agreement with the
rotational temperature  inferred from a  population diagram (\mbox{Fig.~\ref{fig:DRs_corr_c2h}}).

\section{C$_2$H rotational temperatures
and $N$(C$_2$H) toward the  SDLS position} 
\label{sec:rot-diag}

To understand the excitation conditions that lead
to the observed (sub)mm C$_2$H rotational  emission, we re-analyzed
the C$_2$H \mbox{$N$\,=\,1--0} to \mbox{4--3} \citep[IRAM\,30\,m telescope,][]{Cuadrado15} and \mbox{$N$\,=\,6--5} to 
\mbox{10--9} \citep[Herschel/HIFI,][]{Nagy15,Nagy17} lines detected
toward the single-dish line-survey (SDLS) position (see \mbox{Fig.~\ref{fig:RGB_Bar} left}).
The angular resolution of these single-dish observations varies
with frequency, from $\sim$8$''$ to 42$''$, which is not sufficient to spatially resolve the emission from the main small-scale DFs
(a region of $\Delta(\delta x)$\,$\simeq$\,10$''$ width). 
We corrected the observed (single-dish) line intensities toward the SDLS position
 by the beam dilution factors estimated in Appendix~\ref{App:beam_dilution} from our IRAM\,30\,m and ALMA 
C$_2$H \mbox{$N$\,=\,4--3} maps. This method assumes that all rotational lines have the
same spatial distribution. Table~\ref{table:f_b} summarizes the correction factor ($f_b$) applied to each  line. In \mbox{Sect.~\ref{sec:pdr_mods}} we used the corrected line intensities to determine the best 
thermal pressure value, $P_{\rm th}$/$k_B$, from isobaric PDR models  
 (see Fig.~\ref{fig:PDR2MTC_C2H} to compare different
model results).

\begin{figure}[h]
\centering   
\includegraphics[scale=0.67, angle=0]{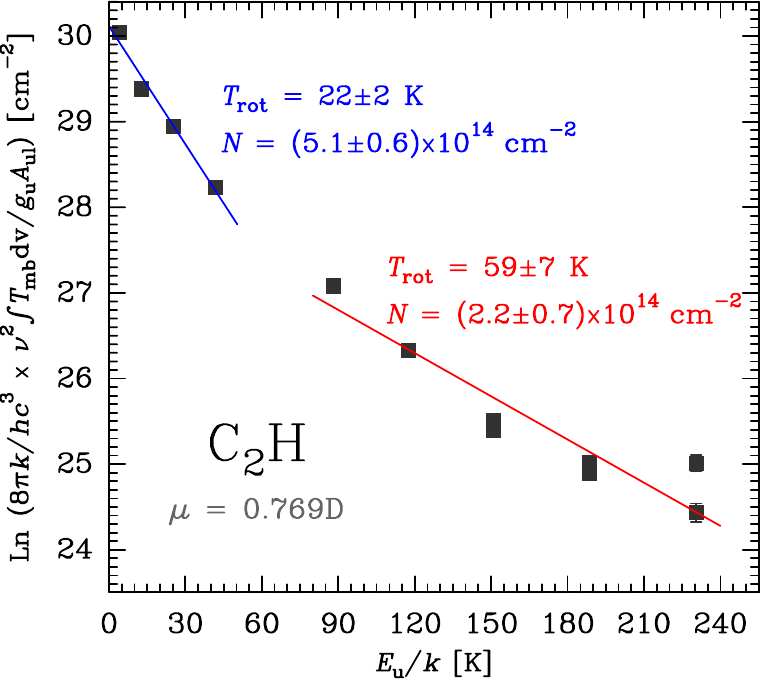}
\caption{Rotational diagram of C$_2$H
determined from  IRAM\,30\,m \citep{Cuadrado15} and Herschel/HIFI \citep{Nagy17} 
observations toward
the  SDLS position in the Bar.   We corrected the line intensities with
the frequency-dependent beam-coupling factors ($f_{\rm b}$) calculated in 
 Sect.~\ref{App:beam_dilution}. 
This diagram shows the fitted column density $N$(C$_2$H), rotational temperature
$T_{\rm rot}$, and their uncertainties.} 
\label{fig:DRs_corr_c2h}
\end{figure}

With the corrected line intensities, we  constructed a rotational population diagram.  The resulting diagram (Fig.~\ref{fig:DRs_corr_c2h}) can be fitted with two rotational temperature components, \mbox{22\,$\pm$\,2~K} and \mbox{59\,$\pm$\,7~K}, and  \mbox{$N$(C$_2$H)\,=\,(7.3\,$\pm$1.3)\,$\times$\,10$^{14}$\,cm$^{-2}$} (assuming 
\mbox{Boltzmann} populations and optically thin line emission).
These values illustrate the excitation conditions and  $N$(C$_2$H) in
the SDLS position.
The positive curvature of the population diagram can also
reflect either a single-gas-temperature, very subthermal component
(with \mbox{$T_{\rm rot}$(C$_2$H)\,$\ll$\,$T_{\rm 64}$(H$_2$)\,$\simeq$\,$T_{\rm k}$}) 
or  a distribution of gas temperatures \citep[e.g.,][]{Neufeld12}  enclosed in the large beam of these single-dish observations, perhaps more consistent with the sharp temperature gradients in the  PDR.

  
\section{P$_{\rm th}$ determination from C$_2$H lines}
\label{subsec:pdr2mtc}

Here we compare the complete set of (beam-dilution corrected) \mbox{C$_2$H $N$\,=\,1--0} to \mbox{10--9}  intensities  observed toward the SDLS position  (including DF3 and DF2; \mbox{Fig.~\ref{fig:RGB_Bar}}) with isobaric PDR models of different $P_{\rm th}$/$k_{\rm B}$ values.  
We used the output of different models: $T_{\rm k}$, $T_{\rm d}$, $n_{\rm H}$, and $n$(C$_2$H)  profiles from $A_V$\,=\,0 to 10\,mag, as input for a    nonlocal 
C$_2$H excitation and 
radiative transfer  calculation.

\begin{figure}[t]
\centering   
\includegraphics[scale=0.43, angle=0]{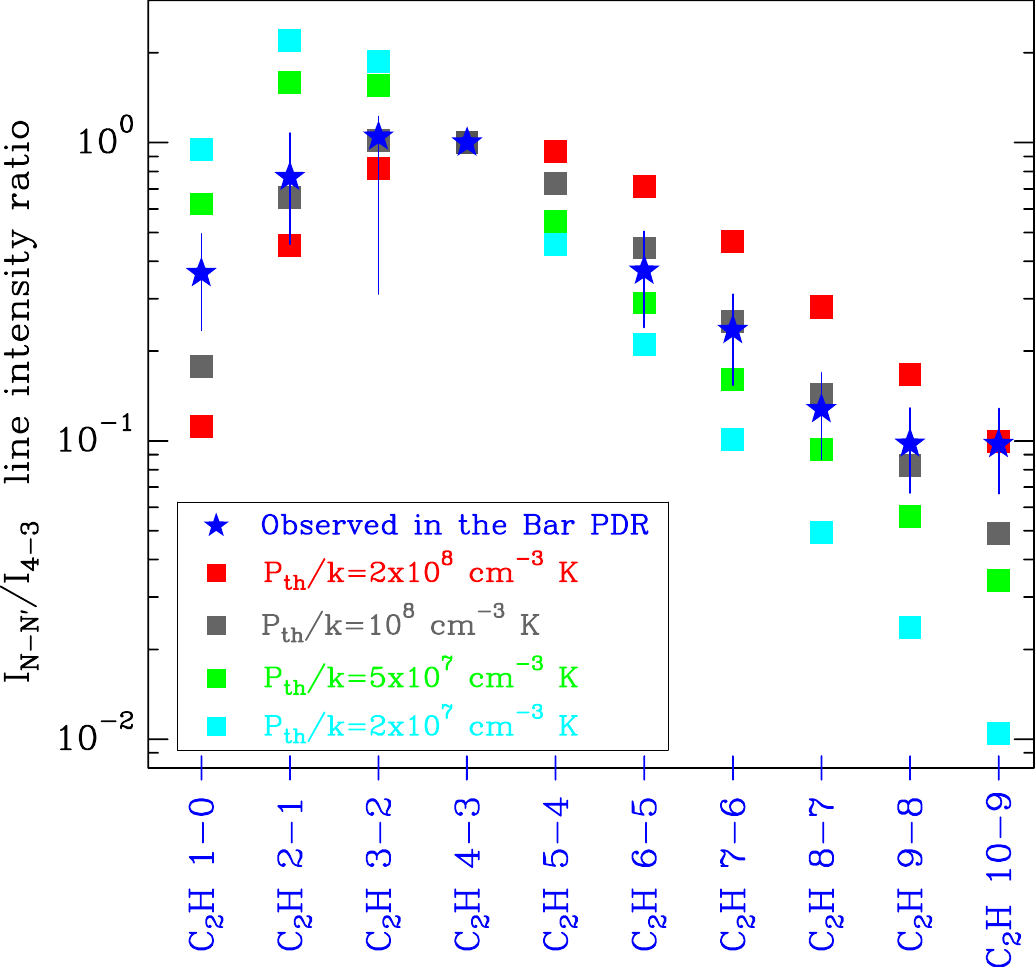}
\caption{Normalized C$_2$H line intensities predicted by  isobaric PDR models of different $P_{\rm th}$/$k_{\rm B}$. 
Blue stars represent the observed intensities, corrected by beam-dilution, toward the SDLS position of the Bar.
Vertical lines represent the uncertainty of the observational  intensity ratios.}
\label{fig:PDR2MTC_C2H}
\end{figure}

The C$_2$H line intensities predicted by PDR models with \mbox{$P_{\rm th}$/$k_{\rm B}$\,=\,(0.5--1)$\times$10$^8$\,K\,cm$^{-3}$} bracket the observed line intensities.  
The gas thermal pressure and $G_0$ control the chemistry at the DF, through the production of \mbox{H$_{2}^{*}$($v$$\geq$1)} (which triggers the formation of simple hydrocarbons,  see \mbox{Sect.~\ref{subsec:gas_chemistry}}). \mbox{$P_{\rm th}$/$k_{\rm B}$} sets the gas density (which drives the collisional excitation of the observed  lines). 
Since the Bar is not a perfectly \mbox{edge-on} PDR, comparing absolute line intensities requires knowledge of the tilt angle $\alpha$  relative to a purely \mbox{edge-on} PDR. Thus, we compare the C$_2$H line intensities normalized by the 
\mbox{C$_2$H~$N$\,=\,4--3} line intensity. As these rotationally excited lines are effectively optically thin (i.e., their intensities are proportional
to $N$(C$_2$H)), the assumption of a tilt angle leads to the same geometrical intensity enhancement factor for all lines. 
 \mbox{Figure~\ref{fig:PDR2MTC_C2H}} shows the predicted normalized   \mbox{$I$(C$_2$H $N-N'$)/$I$(C$_2$H $4-3$)} line intensities for different $P_{\rm th}$ values (colored squares) along with the observed line intensity ratios (blue stars). 
 Models with  \mbox{$P_{\rm th}$/$k_{\rm B}$\,=\,2$\times$10$^8$\,K\,cm$^{-3}$}
produce peak abundances of [C$_2$H]\,$\simeq$\,10$^{-6}$ and
gas densities that are too high. They overestimate the observed   \mbox{high-$N$/4--3}   intensity ratios. On the other hand, models with the lowest pressure lead to much lower gas densities (several 10$^4$\,cm$^{-3}$) and low
[C$_2$H] peak abundances, a few 10$^{-8}$. This leads to  
 faint C$_2$H emission and very low \mbox{high-$N$/4--3}  intensity ratios.
The best agreement is for  models with
 \mbox{$P_{\rm th}$/$k_{\rm B}$\,$\simeq$\,10$^8$\,K\,cm$^{-3}$}. This should be taken as the typical pressure in the bright C$_2$H--emitting layers. 
 The poorer agreement with the observed
 \mbox{$I$(C$_2$H $10-9$)/$I$(C$_2$H $4-3$)} ratio may suggest either higher density  substructures exist within the large beam enclosed by the single-dish observations \mbox{\citep[e.g.,][]{Nagy15,Andree17}} or additional excitation mechanisms, 
 such as 
 chemical formation pumping, 
that play a role in the excitation of C$_2$H  excited levels.

\section{Fine-structure excitation of C($^3$P) by H$_2$}\label{App:C_rates}

\cite{Klos_18,Klos_18_corr} provided rate coefficients for the fine structure excitation of C($^3$P) by inelastic collisions with H$_2$ in the 10-100~K range. The scattering calculations were based on highly correlated C($^3$P)--H$_2$ potential energy surfaces (PESs)  computed at the explicitly correlated multireference configuration interaction level of theory \citep{Shiozaki:11} using a large atomic basis set. Quantum scattering close-coupling equations were solved using the HIBRIDON package \citep{hibridon5} in order to get the inelastic cross sections and rate coefficients. The spin-orbit energy levels of C($^3$P$_J$) atom are 0, 16.41671 and 43.41350 cm$^{-1}$ for $J = 0,1$ and 2, respectively. Details about the calculations can found in \cite{Klos_18,Klos_18_corr}.

The scattering data were compared in great detail with experimental measurement at low collisional energies \citep{Klos_18,Klos_18_corr,Plomp:23} and a very good agreement was found. This comparison validates the high accuracy of the C($^3$P)--H$_2$ PES developed and of the scattering approach used. 

\begin{figure}[h]
\centering   
\includegraphics[scale=0.57, angle=0]{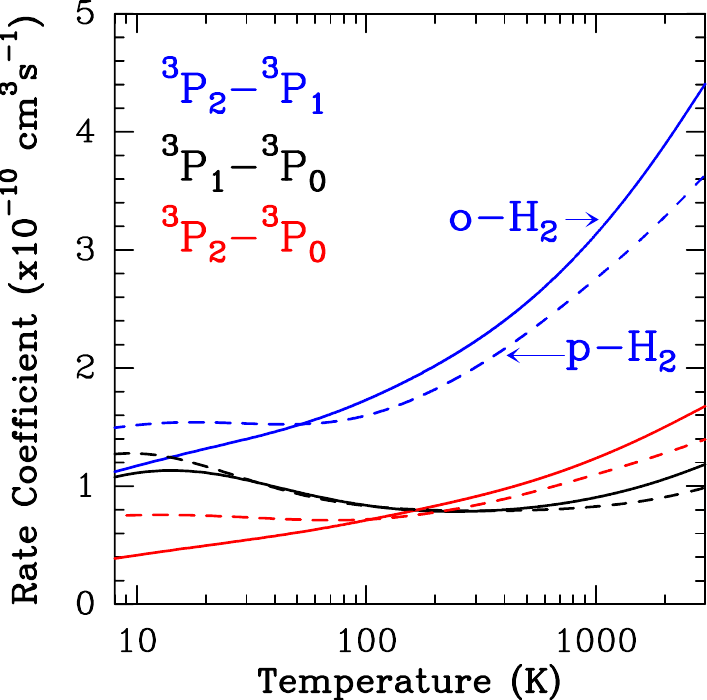}
\caption{Rate coefficients for inelastic collisions between atomic carbon
C($^3$P) and \mbox{$o$-H$_2$($J$=1)} (continuous curves) and \mbox{$p$-H$_2$($J$=0)}
(dashed).}
\label{fig:C_rates}
\end{figure}

Here, we extend the scattering calculations to higher temperatures (up to 3000~K) in order to cover the astrophysical needs for modeling PDRs and the warm neutral  medium. Scattering calculations were performed for collision energies up to \mbox{15,000 cm$^{-1}$}. The close-coupling equations are propagated using hybrid Alexander-Manolopoulos propagator from the initial distance of $R = 1.0$ $a_0$ to 80 $a_0$, $R$ being the distances between C and the center of mass of H$_2$. The cross sections were checked for convergence with respect to the inclusion of a sufficient number of partial waves and energetically closed channels. The H$_2$ basis included all levels with a rotational quantum number $J \le 6$ belonging to the ground vibrational state manifold, and the contributions of the first  200 partial waves were included in the calculations at 15,000 cm$^{-1}$. Thermal rate coefficients in \mbox{cm$^3$\,s$^{-1}$} 
(\mbox{Fig~\ref{fig:C_rates}}) were obtained by an integration of the cross sections over a Maxwell-Boltzmann distribution of the collisional energy.

We compared the new rate coefficients to those computed by \cite{Schroder:11} 
and frequently used in astrophysical models. The global agreement is good even if deviations of $\sim$20\% exist at low $T$. 
The differences can be explained by the use of different PESs and couplings between electronic states.

\end{appendix}

\end{document}